\documentclass[fleqn,usenatbib]{mnras}
\usepackage{newtxtext,newtxmath}
\usepackage{amsmath}
\usepackage{hyperref}
\usepackage[T1]{fontenc}
\usepackage[utf8]{inputenc}
\usepackage[pdftex]{graphicx}
\usepackage{amsmath}
\usepackage{float}
\usepackage[dvipsnames,table,xcdraw]{xcolor}

\usepackage[english,nameinlink,noabbrev]{cleveref}
\usepackage{diagbox}
\usepackage{multirow}
\newcommand{\bs}[1]{\mathbf{#1}}
\newcommand{\mc}[1]{\mathcal{#1}}
\newcommand{\dif}{\mathrm{d}} 
\newcommand{\ts}[1]{\textsc{#1}}

\newcommand{\mcov}{\boldsymbol{\mathsf {C}}}
\newcommand{\mcor}{\boldsymbol{\mathsf {R}}}

\title[LIGER II]{The large-scale monopole of the power spectrum in a Euclid-like survey: wide-angle effects, lensing, and the `finger of the observer'}

\author[Elkhashab, Porciani \& Bertacca]{
Mohamed Yousry Elkhashab,$^{1,2,3}$\thanks{E-mail:elkhashab@pd.infn.it}
Cristiano Porciani ,$^{2}$
and Daniele Bertacca $^{1,3,4}$
\\
$^{1}$Dipartimento di Fisica e Astronomia Galileo Galilei Universita‘ di Padova, 35131 Padova,
Italy\\
$^{2}$Argelander-Institut für Astronomie, Auf dem Hügel 71, D-53121 Bonn, Germany\\
$^{3}$ INFN Sezione di Padova,  I-35131 Padova, Italy.\\
$^{4}$INAF - Osservatorio Astronomico di Padova, Vicolo dell'Osservatorio 5, I-35122 Padova, Italy.
}

\date{Accepted 2021 October 12. Received 2021 October 11; in original form 2021 August 12}

\pubyear{2021}
\begin{document}
\label{firstpage}
\pagerange{\pageref{firstpage}--\pageref{lastpage}}
\maketitle

\begin{abstract}
Radial redshift-space distortions due to peculiar velocities and other light-cone effects shape the maps we build of the Universe.  We address the open question of their impact onto the monopole moment of the galaxy power spectrum, $P_0(k)$. Specifically, we use an upgraded numerical implementation of the \ts{liger} method to generate $140$ mock galaxy density fields for a full Euclid-like survey and we measure $P_0(k)$ in each of them utilising a standard estimator. We compare the spectra obtained by
turning on and off different effects.
Our results show that wide-angle effects due to radial peculiar velocities generate excess power 
above the level expected within the plane-parallel approximation. They are detectable with a signal-to-noise ratio of 2.7 for $k<0.02\,h$ Mpc$^{-1}$.  Weak-lensing magnification also produces additional power on large scales which, if the current favourite model for the luminosity function of H$\alpha$ emitters turns out to be realistic, can only be detected with a signal-to-noise ratio of 1.3 at best.
Finally, we demonstrate that measuring $P_0(k)$
in the standard of rest of the observer
generates an additive component
{reflecting the kinematic dipole overdensity caused by the peculiar velocity. This component is}
characterised by a damped oscillatory pattern on large scales.  We show that this `finger of the observer' effect is detectable in some redshift bins and suggest that its measurement could possibly open new research directions in connection with the
determination of the cosmological parameters, the properties of the galaxy population under study, and the dipole itself.
\end{abstract}




\section{Introduction}
In order to study the large-scale structure of the Universe,
we measure redshifts and angular positions of large samples of galaxies and use the information to build three-dimensional maps of
the galaxy distribution. This mapping is done 
by assuming that the light bundles emitted by the galaxies propagate in an unperturbed Friedmann-Lema\^{i}tre-Robertson-Walker (FLRW) model universe.
Neglecting the rich complex structure that lies in between us
and the distant galaxies gives rise to the so-called 'redshift-space distortions' (RSD) in the galaxy maps. 

Galaxy peculiar velocities alter the observed redshifts and thus
generate RSD.
In a seminal paper, \citet{Kaiser87} showed that this
contribution assumes a particularly simple form in Fourier space 
under two assumptions: (i) linear perturbation theory can be used 
and (ii) the galaxies are assumed to fill a relatively small volume lying so far
away from the observer that the radial displacements induced by peculiar velocities can be considered effectively parallel (this is now known as the `distant-observer', '(global) plane-parallel' or `small-angle' approximation).
In this case, the galaxy power spectrum is enhanced and acquires a dependence on the directional cosine between the wave vector and the line of sight.

Relaxing the plane-parallel approximation leads to several complications. The loss of translational invariance induced by radial RSD leads to mode-mode coupling and questions the very
definition of Fourier-based statistics like the power spectrum  \citep{1996ApJ...462...25Z, szalay+98,Hamilton_review}. 
It is thus convenient to introduce the hybrid concept of 
a position-dependent power spectrum that can then be
expanded in Legendre polynomials while averaging over the position
dependence \citep{Scoccimarro2015,REIMBERG2016}.
With the exception of the monopole moment, however, some tricks need to be implemented in order to build fast estimators based on the
Fast-Fourier-Transform algorithm
\citep{Yamamoto_estim, Scoccimarro2015,Bianchi2015, Hand+17}.
These approximations generate errors in the estimation of the quadrupole and the hexadecapole moments \citep{Castorina:2017inr}.
Alternatively, the galaxy density field can be expanded in terms of eigenfunctions of the Laplacian operator, i.e. products of spherical Bessel functions and spherical harmonics
\citep[e.g.][]{1973ApJ...185..413P, Heavens:1994iq, Heavens&Taylor:1997, Percival:2004fs} or 
using hybrid schemes \citep{Wang:2020wsx}.

The forthcoming generation of redshift surveys will measure galaxy clustering on unprecedentedly large scales in the distant Universe.
The very fact that we only detect events lying on our past light cone impacts the observed clustering signal at such length scales. 
These light-cone effects can be computed perturbatively
\citep{Yoo:2009au, Bonvin-Durrer2011, Challinor:2011bk, Jeong:2011as} and include various relativistic corrections that act either locally (i.e. due to peculiar velocities and the Sachs-Wolfe effect) or along the line of sight (e.g. the integrated Sachs-Wolfe effect and the lensing convergence as well as magnification and time delay). On top of this, wide-angle effects 
generate corrections to the plane-parallel approximation for redshift-space distortions \citep{Bertacca:2012tp,Bertacca:2019wyg}. Although these effects can be partially addressed by
expanding any summary statistic (in configuration, Fourier or harmonic space) in terms of the ratio between the pairwise galaxy separations and the comoving distance to the observer  \citep[e.g.][]{Castorina:2017inr}, this approach is doomed to fail when the above ratio is of order unity.

The impact of light-cone and wide-angle effects on the multipole moments of the galaxy power spectrum is an open question in the literature. 
In particular, it is still unclear whether these large-scale corrections (i) can be detected with the next generation of surveys and (ii) can hamper the measurement of primordial non-Gaussianity from the signature of a scale-dependent bias proportional to $k^{-2}$ \citep[e.g.][]{Dalal+08, Verde+08, Giannantonio-Porciani-10}. In this paper, we focus on the first issue. 
To achieve this goal, we 
compute the monopole moment of the observed galaxy power spectrum and its covariance matrix by fully accounting for wide-angle effects and including general-relativistic light-cone corrections to linear order. As a working example, we use a Euclid-like survey \citep{Euclid-Red-Book} but our results can be easily generalised to other future facilities.
In order to address the intricacies of Fourier-based statistics
without resorting to the plane-parallel approximation,
we follow a numerical approach. Specifically, we first use 
the \ts{liger} method \citep{liger}
to build a large suite of mock light cones (covering either the full sky or a survey footprint) that we then utilize to
estimate the monopole moment of the power spectrum.
\ts{liger} combines low-resolution N-body simulations and a perturbative scheme for galaxy biasing and general-relativistic RSD in order to generate galaxy density fields in redshift space.
Since several physical effects
can be turned on and off at will, \ts{liger} is an ideal tool to 
quantify the impact of the different competing processes on any summary statistic.
Our method also allows us to fully account for the observational window function which reflects the finite depth and the incomplete sky coverage of a survey.

We devote special attention to the kinematic dipole overdensity generated in redshift space by our peculiar velocity. 
We jocularly refer to this phenomenon as the impression of the 
`finger of the observer' on a galaxy survey.
Attempts to measure the kinematic dipole from the projected galaxy distribution on the celestial sphere
are hampered by the so-called `local-structure dipole', i.e. by the fact that, for relatively shallow surveys,  the statistical fluctuations in the projected structure at multipole $\ell=1$ are always much larger that the kinematically induced one \citep[e.g.][]{gibelyou-huterer-12}.
Deep galaxy samples extending to redshifts $z>1$ and covering large fractions of the sky are necessary to mitigate this problem \citep[e.g.][]{yoon-huterer-18}.
The impact of the kinematic dipole on 
galaxy-clustering statistics has been mostly neglected so far.
A few theoretical studies exist that
discuss how the dipole influences 
the galaxy two-point correlation function
\citep{Hamilton-Culhane} also accounting
for relativistic light-cone effects
\citep{scaccabarozzi+, Maartens+2018, Bertacca:2019wyg}.
In this work,
we derive an exact analytical expression for the power spectrum of the kinematic dipole  and show that it leads to an enhanced
monopole moment on large scales.
From the analysis of our mock catalogues, we conclude that the forthcoming generation of 
redshift surveys should be able to detect
the additional clustering due to the dipole term.
The amplitude of the correction 
includes contributions from
the redshift dependence of the galaxy
number density \citep{Kaiser87}
but also on 
the expansion history of the Universe
and the magnification bias for magnitude-limited surveys
\citep{Maartens+2018, Bertacca:2019wyg}.
Based on this, we speculate about strategies to further develop the area in the coming years and exploit the kinematic dipole for cosmological inference.

The paper is structured as follows.  After
summarising the main concepts behind the \textsc{liger} method,
in section~\ref{sec:liger}, we describe a new code implementation that improves upon the original one and present a validation test.
Here, we also introduce all our simulations and give a detailed account of the assumptions we adopt to construct the
mock light cones. The monopole moment of the power spectrum
and the estimator we use to measure it from the mock catalogues are introduced in section~\ref{sec:methods}.
Our main results are presented in section~\ref{sec:results} where we discuss how several physical effects alter the monopole of the power spectrum. Furthermore, we use the likelihood-ratio test to assess
which effects can be detected using a Euclid-like survey.
Finally, a summary of our results is laid out in section~\ref{sec:summ}.

\section{LIGER}\label{sec:liger}
\textsc{Liger} \citep[light cones with general relativity,][hereafter paper I]{liger} is a numerical technique for building mock
realisations of the galaxy distribution on the past light cone of an observer. 
By post-processing the snapshots of Newtonian simulation, this method outputs the galaxy distribution in redshift space accounting for relativistic corrections at linear order in the
cosmological perturbations. In this section, we introduce an upgraded code implementation of the \textsc{liger} method and validate it against the popular 
Cosmic Linear Anisotropy Solving System \citep[\textsc{class},][]{CLASSI,CLASSII}.

\subsection{Updates and improvements}
The location of a galaxy in redshift space is obtained by `naively' 
converting its redshift, $z$, and observed position in the sky, $\hat{\mathbf{n}}$, 
under the assumption we live in an unperturbed FLRW
universe. 
The redshift-space position differs from the actual (`real-space') one due to the presence of inhomogeneities which modify both $z$ and $\hat{\mathbf{n}}$ (as well as the observed flux in any waveband). \textsc{liger} is a tool for converting the output of a Newtonian simulation (N-body or hydrodynamic) into a mock realisation of
the past light cone of an observer.
Schematically, \textsc{liger} builds a mapping between real- and redshift space by first shifting the world lines of the synthetic galaxies (in 4-dimensional spacetime) and then determining their intersection with the unperturbed backward light cone of the observer. The first step takes into account both local terms and contributions that have been integrated along the line of sight using the Born approximation (see section 2 in paper I for equations and further details).

This work is based on a numerical implementation of the \textsc{liger} method which
improves on the previous one (paper I) in several different ways.
In terms of additional physics, we take into account the peculiar velocity and potential of the observer in the calculation of the modified worldlines and magnifications, i.e. the first term on the rhs in equations (8), (9), (10) and (12) of paper I (see also appendix~\ref{app:shift_eqs} here for a concise summary). These contributions had been neglected in the original code. A second difference is that we now associate the observer with a simulation particle (the closest one to the user-selected location). From the numerical point of view, we made changes that led to a considerable speed up (especially for the calculation of the line-of-sight integrals) and a reduced memory consumption. Finally, handling I/O operations
has been improved with some completely new features. In its basic form, \textsc{liger} works at the level of
individual galaxies (extracted from hydro simulations or semi-analytic models) which are mapped one by one to redshift space. However, a second implementation of the method has been developed to deal with N-body simulations that cover large comoving volumes but do not resolve individual galaxies. In this case, the \textsc{liger} shift is applied
to the dark-matter particles and a weighting scheme is used to account 
for galaxy biasing and gravitational lensing in a consistent way
(see appendix \ref{galdens} here and section 2.3 of paper I).
This step requires specifying a few statistical properties of the observed galaxy population, namely, the mean number counts $\bar{n}_\text{g}(z)$, the linear-bias parameter $b(z)$, the evolution-bias parameter, $\mc{E}(z)$, and the magnification-bias parameter $Q(z)$, all as a function of redshift. Once the dark-matter density on the past light cone has been computed, it is straightforward to generate a mock galaxy density field based on these functions. In order to facilitate the usage of \textsc{liger} in this `large-box mode', we have developed the \textsc{buildcone} toolkit which we make publicly available together with the upgraded \textsc{liger} code
at this \href{https://astro.uni-bonn.de/\~porciani/LIGER/}{URL}. \textsc{buildcone} reads the output of \textsc{liger} (i.e. the lightcone positions of the dark-matter particles in real and redshift space) and outputs the (gridded) galaxy density
field in redshift space.

\subsection{Simulations}
\label{sec:sims}
In this paper, we quantify the impact of light-cone and wide-angle effects on the galaxy power spectrum monopole on very large scales.
As a reference case,
we present results for an Euclid-like spectroscopic survey in the redshift range $0.9<z<1.8$.
This corresponds to a maximum comoving radial distance $x_{\mathrm{max}}\simeq 5$ Gpc $= 3.4\, h^{-1}$ Gpc. 
Therefore, we need to simulate the matter density field in boxes with a comoving side of at least $2x_{\mathrm{max}}$.
In order to account for density perturbations with wavelengths larger than the survey footprint, we actually use cubic periodic boxes with a comoving side of $L=12\,h^{-1}$ Gpc.
Since we are interested only in the power-spectrum monopole on linear and quasi-linear scales, 
to minimize the computational costs, we run a suite of 35 simulations based on second-order Lagrangian perturbation theory (\ts{2LPT})
using the \textsc{music} code \citep{music}.
We use $1024^3$ particles with a mass of $m_{\mathrm{p}}=1.4\times 10^{14}\,h^{-1} \mathrm{M}_{\sun}$
that are initially forming a regular Cartesian grid with a linear spacing of $\Delta x=11.7\,h^{-1}$ Mpc.
Note that the particle distribution forms a slightly perturbed grid at all redshifts of interest and no shot noise correction should be applied when the power spectrum is computed
\citep{Gabrielli04, DiscretenessIC07,AccurateP08}.

We adopt a basic $\Lambda$CDM model with matter density parameter
$\Omega_\mathrm{m}=0.3158$, baryon density parameter $\Omega_{\mathrm{b}}=0.0508$, present-day dimensionless Hubble constant $h=0.673$ as well as primordial scalar perturbations with spectral index $n_{\mathrm{s}}=0.966$ and amplitude
$\sigma_8=0.812$ \citep{planck18}.

\begin{figure}                                              \includegraphics[width=1\columnwidth, keepaspectratio=true]{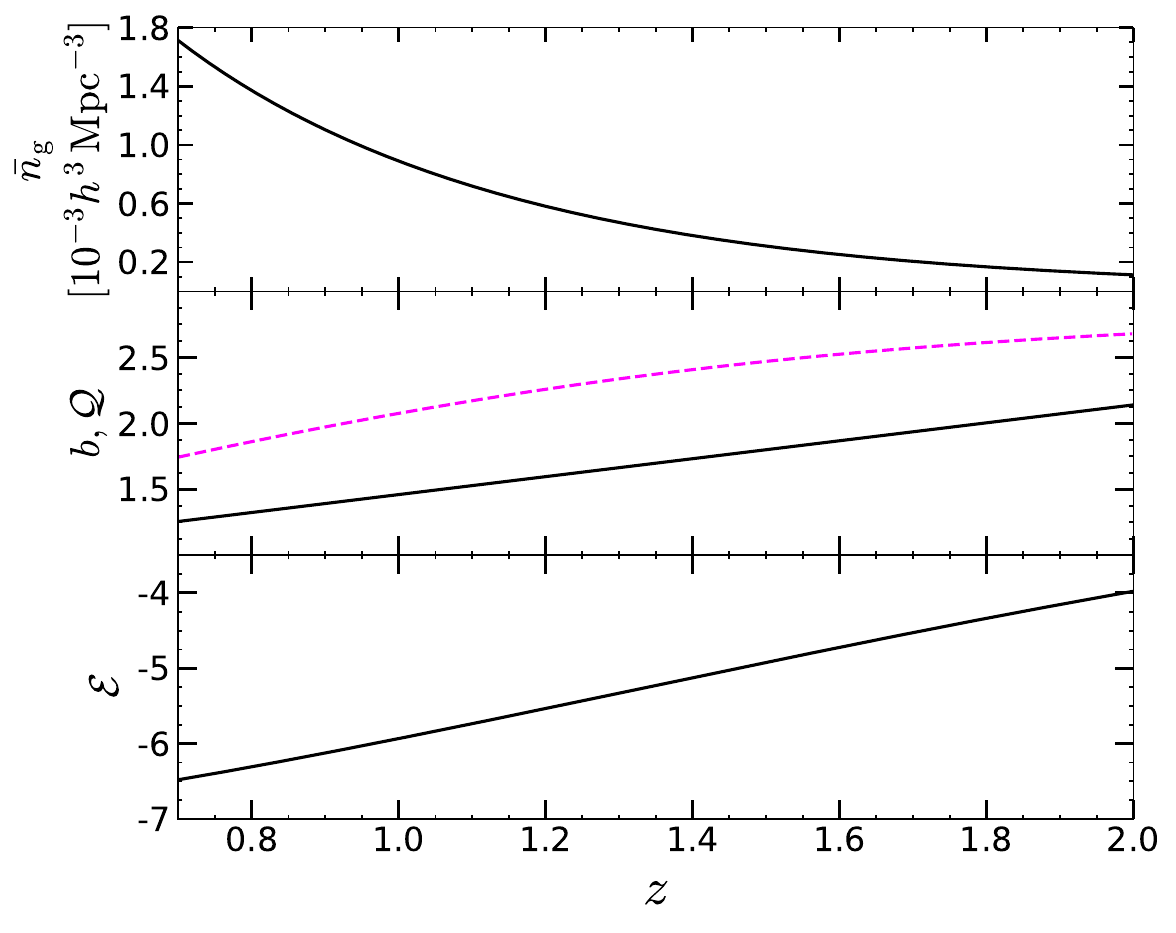}
\caption{The top panel shows
the mean galaxy number density 
for a Euclid-like spectroscopic survey as a function of redshift.
This quantity is 
computed using model 3 for the H$\alpha$ luminosity
function in \citet{pozzetti16} and assuming a flux limit of $2\times 10^{-16}$ erg cm$^{-2}$ s$^{-1}$ with a completeness of 70 per cent.
The middle panel displays the corresponding magnification bias (magenta dashed line) and the assumed redshift dependence for the linear bias parameter (black solid line). The bottom panel shows the evolution bias.}
\label{Fig:functions}                               
\end{figure}

\subsection{Mock Euclid-like catalogues}
\label{sec:mock}
\begin{figure*}                                         
\includegraphics
[width=1.89\columnwidth, keepaspectratio=true,trim={0.4mm 0.4mm 0.4mm 0.4mm},clip]
{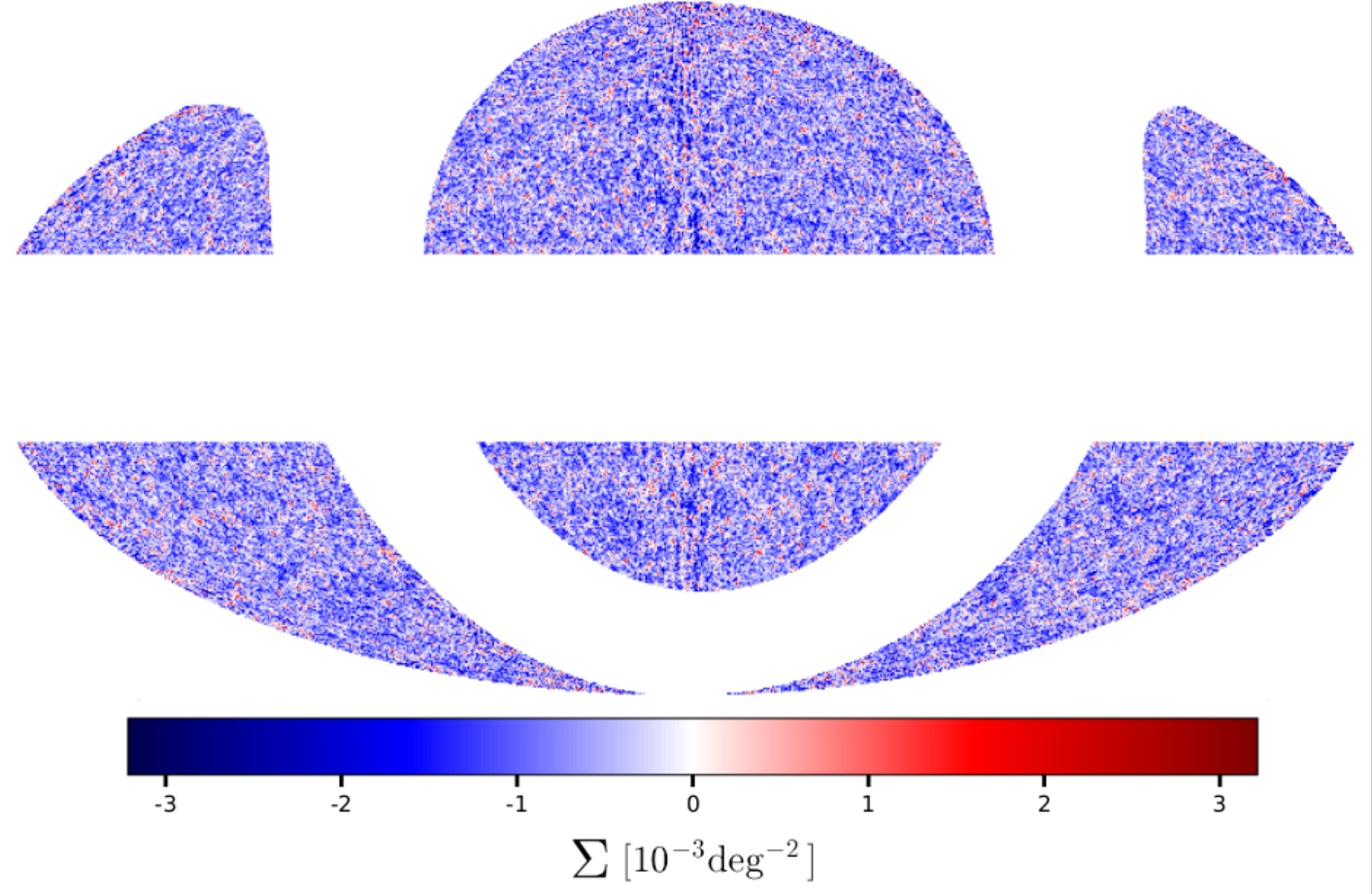}
\caption{Mollweide projection in ecliptic coordinates of the galaxy number density of an Euclid-like survey in the redshift range $0.9<z<1.1$. The survey footprint covers 15,000 square degrees obtained by masking the regions within 20 degrees from both the Galactic and ecliptic planes.}           
\label{fig:proj_euc}                               
\end{figure*}

In order to `paint' galaxies on top of the underlying matter distribution,
the \textsc{liger} method requires assuming a  
model for the redshift-dependent luminosity function and one for the linear bias parameter of the galaxy population of interest.
Regrettably, little information is available
about the H$\alpha$ emitters that will be targeted in the Euclid wide spectroscopic survey.
A recent compilation of emission line galaxies
\citep{bagley20} is in rather good agreement with  the luminosity function described by Model 3 in \citet{pozzetti16} that we then adopt here. This is also one of the models that have been used by the Euclid consortium to build mock catalogues based on the Euclid Flagship Simulation.
Let us denote by $n(L_\textrm{min},z)$ the comoving number density of galaxies with H$\alpha$ luminosity greater than $L_\textrm{min}$ at redshift $z$.
Assuming that the spectroscopic sample is flux limited,
we write its selection function\footnote{Although the selection function of an actual survey might depend also on the sky position or the size of a galaxy, we only consider
radial variations in this work.} (i.e.
the probability that a galaxy within the luminosity range $(L_\textrm{min}, L_\textrm{min}+\dif L_\textrm{min})$ and
the redshift range $(z,z+\dif z)$ is included in the survey) as $\Theta[L_\textrm{min}-L_\text{lim}(z)]$ where $\Theta$ is the Heaviside step distribution function and $L_\text{lim}(z)$ indicates
the redshift-dependent luminosity corresponding to the flux limit.
Given the selection criteria of the survey,
we obtain
the expected mean number density of galaxies with a measured redshift,   $\bar{n}_\text{g}(z)=n(L_\text{lim}(z),z)$,
by integrating the luminosity function above an H$\alpha$ flux limit of
$2\times 10^{-16}$ erg cm$^{-2}$ s$^{-1}$ 
and assuming a completeness of 70 per cent \citep{Euclid-Red-Book}.
We then compute
the evolution-bias parameter,
\begin{equation}
    \mc{E}(z)=-\left.\frac{\partial \ln n(L_\textrm{min},z)}{\partial \ln (1+z)} \right|_{L_\textrm{min}=L_\text{lim}(z)}\;,
    \label{eq:evbias}
\end{equation}
and the magnification-bias parameter,
\begin{equation}
Q(z)=-\left.\frac{\partial \ln n(L_\textrm{min},z)}{\partial \ln L_\textrm{min}} \right|_{L_\textrm{min}=L_\text{lim}(z)}\;,
\label{eq:magbias}
\end{equation}
in a consistent way (see appendix~\ref{sec:evmagbias} for details). For the linear bias, instead, we assume the relation
\begin{equation}
b(z)=1.46+0.68(z-1)\;,
\end{equation}
which has been derived by adapting table 3 in \citet{EuclidVII}. These input functions are displayed in Fig.~\ref{Fig:functions}.

We extract four non-overlapping light cones from each of the simulations described in section~\ref{sec:sims},
for a total of $N_\text{real}=140$ full-sky realisations.
In each of them, 
the galaxy density on the past light cone of the observer, $n_\mathrm{g}(\mathbf{x})$ 
(where $\mathbf{x}$ denotes the comoving position in redshift space), is obtained
using equation~(\ref{eq:master_eq}) in 
appendix~\ref{galdens}. 
In short, $b(z)$, $\mc{E}(z)$, $Q(z)$, and
the magnification ${\mathcal M}(\mathbf{x},z)$ (lensing+Doppler) provide weights
for combining different particle distributions obtained from applying the \ts{liger} code to the \textsc{music} runs. 
As we have already mentioned in section~\ref{sec:sims}, shot noise is negligible in our mock catalogues by construction. However, this will not be the case in an actual survey. For this reason, before discussing the detectability of the different effects with a Euclid-like survey, we add Poissonian shot noise to our light cones  as described in appendix~\ref{sec:shotnoise}.

Practically,
we first use the \ts{buildcone} toolkit  to sample $n_\mathrm{g}(\mathbf{x})$ on a regular Cartesian grid. 
We then mask out the regions that lie within 20 degrees from both the Galactic and the ecliptic planes so that to reproduce the characteristic footprint of an Euclid-like wide spectroscopic survey.
A sample map of the resulting projected galaxy density in the redshift bin $0.9<z<1.1$ is shown in Fig.~\ref{fig:proj_euc}. 

In order to assess the importance of different physical effects, for each of the 140 realisation, we build five different mock light cones which progressively take into account an increasing number of phenomena (see appendix~\ref{app:shift_eqs} for the mathematical details).
In the simplest case, we study
the galaxy distribution in real space (REAL). Next, we consider the redshift-space distortions caused by shifting the simulation particles along the line of sight due to their peculiar velocity (V$_\text{cmb}$).
In most studies, this is the only contribution which is taken into account.
As a third option (GR$_\text{cmb}$), we turn on all linear relativistic effects (gravitional and Doppler lensing, Sachs-Wolfe effects, Shapiro time-delay, etc.)
barring those related to the peculiar velocity of the observer.
This 
corresponds to using galaxy redshifts
relative to the standard of rest in which the cosmic microwave background (CMB) is isotropic.
Finally, we investigate what happens
if we use redshifts defined with respect
to the standard of rest of the observer. In this case, we build two
additional mock light cones. Proceeding as above, in the first, we only consider the radial shift due to the peculiar velocities (V$_\text{obs}$)
while, in the second, we account for all possible linear relativistic effects (GR$_\text{obs}$)

\subsection{Validation}
In order to validate our numerical implementation of \ts{liger},
we compare our results against those obtained with the \ts{class} code.
In particular, we focus on the angular power spectra in real and redshift space that can be easily computed using both codes. Note that
\ts{class} numerically integrates the perturbative expression for
the power spectrum at leading order while \ts{liger} generates mock catalogs from which we estimate the angular power spectrum (as detailed in appendix \ref{ligcell}). The \ts{liger} results are therefore subject to sample and estimation variance. For a meaningful comparison, we average the \ts{liger} power spectra over 140 full-sky realisations. 
\begin{figure*}
\includegraphics[width=1.9\columnwidth, keepaspectratio=true]{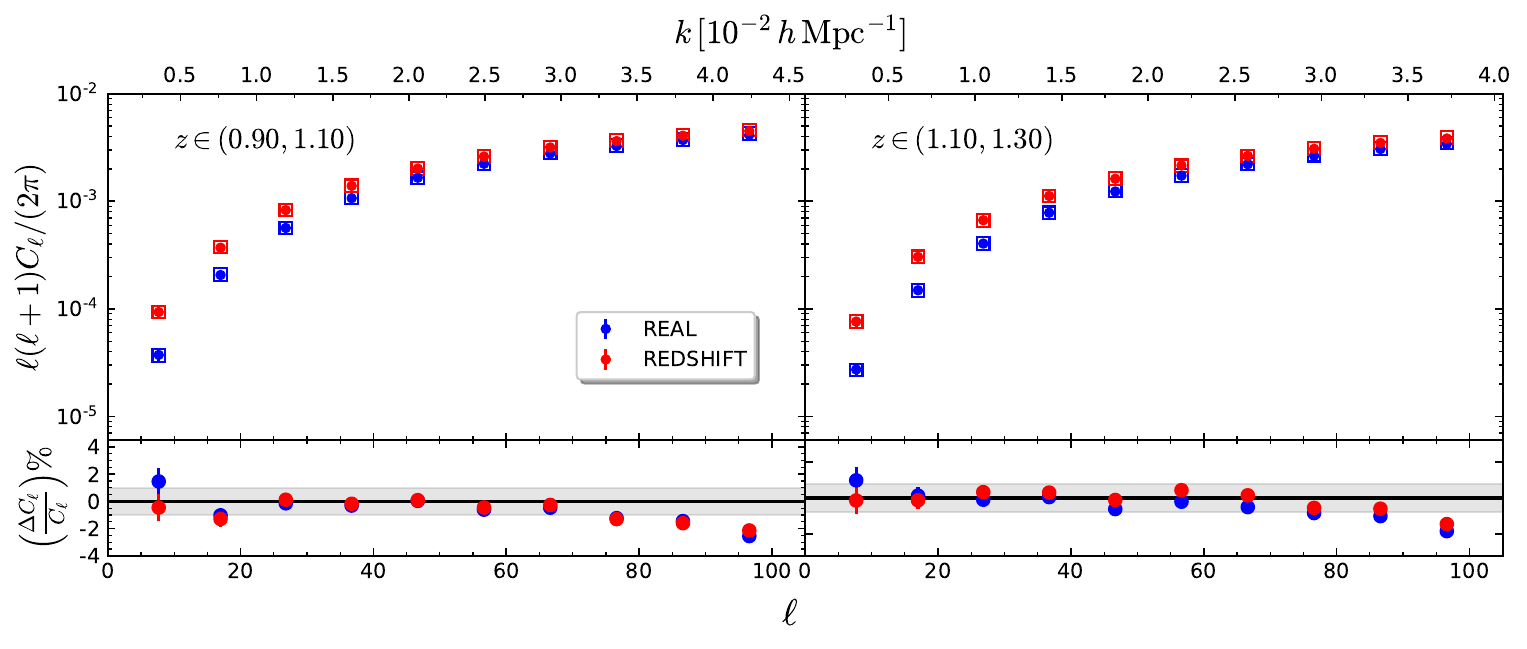}
\caption{Binned angular power spectra ($\Delta \ell=10$) obtained with \ts{class} (empty squares) and from 140 mock catalogues generated with \ts{liger} (solid stars) in real space (blue) and in redshift space including all light-cone effects (red). The bottom panels show the relative difference $\Delta C_\ell/C_\ell=(C_\ell^{\ts{liger}}-C_\ell^{\ts{class}})/C_\ell^{\ts{class}}$. The top axis approximately converts the multipole order $\ell$ into a comoving wavenumber $k$ at the central redshift.}
\label{fig:ANG_SPEC} 
\end{figure*}
For simplicity, in this test case, we assume a galaxy population characterised by\footnote{The magnification-bias parameter, $s$, used in \ts{class} is defined in terms of galaxy magnitudes instead of luminosities. It follows from the definitions that $2Q=5s$. Lensing magnification introduces leading-order corrections to the observed galaxy density contrast which scale as $1-Q$, therefore, by setting $Q=0$, we do not erase them.
}
$b=1$, $Q=0$ and a number density that slightly decreases with redshift  (see the top panel of Fig.~\ref{Fig:functions}). We consider a full-sky survey and partition the celestial sphere into more than 3 million equal area pixels which give a robust estimation of the angular power spectrum up to $\ell\simeq 1500$.
Results for the redshift bins
$0.9<z<1.1$ and $1.1<z<1.3$ 
are shown in Fig.~\ref{fig:ANG_SPEC} where we plot the angular power spectrum $\ell (\ell+1) C_\ell/(2 \pi)$ as a function of the multipole order $\ell<100$
(the top axis approximately converts $\ell$ into a comoving wavenumber using the relation $k\simeq (\ell+1/2)/x(\bar{z})$ where $x(\bar{z})$ denotes the comoving radial distance in the background cosmology and $\bar{z}$ is the central redshift in each bin).
\ts{class} and \ts{Liger} agree to better than one per cent for
$\ell\lesssim 85$ . For smaller angular scales, the \ts{liger} spectra
show slightly less power than those obtained with \ts{class}. 
The reason for this is threefold. First, on these scales and redshifts,
non-linear physics generates a small reduction in power compared to linear predictions \citep[see e.g. figure 7 in][]{wer-por}. However, 
2LPT does not fully capture this physical effect and gives rise to power spectra that slightly differ with respect to actual N-body simulations \citep[see e.g. figure 6 in][]{Taruya+18}. Finally, the rather coarse mesh used in our large computational boxes causes a spurious loss of power as $k$ approaches the Nyquist wavenumber $k_\text{N}=\pi/\Delta x\simeq 0.268\,h$ Mpc$^{-1}$ \citep[see section 3 in][]{music}.
Note that this is a limitation of the simulations we use and not of \ts{liger}. 
Anyway, in this work we will focus on the largest scales probed by the next generation of galaxy redshift surveys ($k<0.02\, h$ Mpc$^{-1}$)
where \ts{class} and \ts{liger} are in spectacular agreement.
\section{Monopole moment of the power spectrum}\label{sec:methods}

In this section, we address the monopole moment of the power spectrum in redshift space and its covariance matrix
by fully accounting for wide-angle effects and including general-relativistic light-cone corrections to linear order.
As a reference case,
we present results for an Euclid-like spectroscopic survey in the redshift range $0.9<z<1.8$.

\subsection{The galaxy power spectrum in real space}
\label{sec:Preal}
We generally assume that the galaxy density contrast  in real space $\delta_\text{g}(\bs{x})$ can be modelled as a statistically homogeneous and isotropic random process with two-point autocorrelation function $\langle
\delta_\text{g}(\bs{x})\,\delta_\text{g}(\bs{x}+\bs{r})\rangle=\xi(r)$ (where the brackets denote averaging over an ensemble of realisations).

For any absolutely integrable function, $f(\bs{x})$,
we can introduce the Fourier transform 
$\tilde{f}(\bs{k}) = \int\, f(\bs{x})\,e^{-i\mathbf{k}\cdot\mathbf{x}}\, \dif^3x$.
This definition, however, cannot be directly applied to   a (stationary) random field over the whole $\mathbb{R}^3$ such as  $\delta_\text{g}(\bs{x})$   as,
in most realisations, $\int |\delta_\text{g}(\bs{x})|\,\dif^3x$ diverges. 
Let us thus consider a
finite compact region of space $\mathcal{V}$ with volume $V$ and introduce the function $\delta_\mc{V}(\bs{x})=\delta_\text{g}(\bs{x})\,I_{\mathcal V}(\bs{x})$ where
$I_{\mathcal V}$ is the indicator (or characteristic) function of $\mc{V}$ (which is equal to one if $\mathbf{x}\in{\mathcal V}$ and zero otherwise).
We define the power spectral density (or, simply, the power spectrum) of $\delta_\text{g}$ as

\begin{equation}
    P(k)=\lim_{V\to \infty}\frac{ \langle \left| \tilde{\delta}_\mc{V}(\mathbf{k})\right|^2\rangle}{V}=\lim_{V\to \infty}\frac{ \langle \tilde{\delta}_\mc{V}(\mathbf{k})\, \tilde{\delta}_\mc{V}(-\mathbf{k})\rangle}{V}\;,
    \label{eq:PdefV}
\end{equation}
where we have used the fact that,
since $\delta_\text{g}(\bs{x})\in \mathbb{R}$,
then $\tilde{\delta}_\text{g}^*(\bs{k})=
\tilde{\delta}_\text{g}(-\bs{k})$.
Note that the limit in equation~(\ref{eq:PdefV}) exists only if it is taken after averaging over the ensemble.
It follows from the definitions above that
\begin{equation}
\langle \tilde{\delta}_\mc{V}(\mathbf{k}_1)\,\tilde{\delta}_\mc{V}(\mathbf{k}_2)\rangle=\int \xi(r)\,e^{-i\bs{k}_1\cdot\bs{r}}\,\dif^3r\int_\mc{V} e^{-i(\bs{k}_1+\bs{k}_2)\cdot \bs{x}}\,\dif^3x\;,
\end{equation}
where the first integral is taken over the separation 
vectors $\bs{r}=\bs{x}_2-\bs{x}_1$ such that $(\bs{x}_1,\bs{x}_2) \in  \mc{V}\times\mc{V}$.
By taking the limit $V\to \infty$ and extending the
definitions to generalised functions, we obtain
\begin{equation}
\lim_{V\to \infty}    \langle \tilde{\delta}_\mc{V}(\mathbf{k}_1)\,\tilde{\delta}_\mc{V}(\mathbf{k}_2)\rangle=(2 \pi)^3\,P(k)\,\delta_\text{D}(\bs{k}_1+\bs{k}_2)\;,
\end{equation}
where $\delta_\text{D}(\bs{x})$ denotes the three-dimensional Dirac delta distribution.
This result
shows that the power spectral density and the 2-point
autocorrelation function form a Fourier pair (Wiener-Khintchine theorem).
In the remainder of this paper, we will adopt the compact notation
\begin{equation}
\langle \tilde{\delta}_\text{g}(\mathbf{k}_1)\,\tilde{\delta}_\text{g}(\mathbf{k}_2)\rangle=(2 \pi)^3\,P(k)\,\delta_\text{D}(\bs{k}_1+\bs{k}_2)\;,
\label{eq:WKT}
\end{equation}
which is commonly used in the literature.

An estimator
for the  power spectrum can be easily constructed by replacing the
ensemble average $\langle \dots \rangle$ with the
average over a thin shell $V_k$ of Fourier modes in a single realisation such that all
wavenumbers lie in a narrow bin centred around $k$, 

\begin{equation}
    \hat{P}(k)= \frac{1}{V}\,
    \frac{\int_{V_k} \left| \tilde{\delta}_{\mc{V}}(\mathbf{k})\right|^2\,\dif^3k }{\int_{V_k}\mathrm{d}^3k}
    \;.
\end{equation}

\subsection{Definition of the monopole moment in redshift space}
Due to the privileged position of the observer, radial redshift-space distortions break the translational symmetry of the galaxy two-point autocorrelation function. 
The position of a galaxy in redshift space
depends on the difference between its peculiar
velocity and that of the observer. Therefore, the resulting galaxy
overdensity field is not (statistically) homogeneous. On the other hand, 
(statistical) rotational invariance around the observer is preserved.
All this implies that the 2-point autocorrelation function $\langle
\delta_\text{g}(\bs{x}-\bs{r}/2)\,\delta_\text{g}(\bs{x}+\bs{r}/2)\rangle=\xi^{(s)}(\bs{x},\bs{r})$ only depends on three variables: $x$, $r$ and $\hat{\bs{x}}\cdot\hat{\bs{r}}$ (i.e. the cosine of the angle between the galaxy separation and the mean\footnote{
Some authors prefer to use the angle bisector at the observer in order to analytically describe the wide-angle correlation function in configuration space \citep[e.g.][]{szalay+98,Matsubara:1999du,Bertacca:2012tp}.}
line of sight).

The loss of translational invariance
complicates the definition of Fourier-transform-based statistics. For instance, the ensemble average
of the product of two Fourier modes of the overdensity field is not diagonal, i.e. 
\begin{equation}
\langle \tilde{\delta}_\text{g}(\mathbf{k}_1)\,\tilde{\delta}_\text{g}(\mathbf{k}_2)\rangle=(2\pi)^3\,C(\mathbf{k}_1,\mathbf{k}_2) \end{equation}
\citep{1996ApJ...462...25Z,szalay+98}. A position-dependent power spectrum
can be defined as the Fourier transform of the local contribution at $\mathbf{x}$ to the 2-point correlation function at separation $\mathbf{r}$
\citep{Scoccimarro2015}
\begin{equation}
    P_\text{loc}(\mathbf{x},\mathbf{k})=\int \langle \delta_\text{g}\left(\mathbf{x}-\frac{\mathbf{r}}{2}\right)\,\delta_\text{g}\left(\mathbf{x}+\frac{\mathbf{r}}{2}\right)\rangle\,e^{-i\mathbf{k}\cdot\mathbf{r}}\,\mathrm{d}^3r\;.
\end{equation}
Expressing the galaxy overdensity in terms of its Fourier
transform, we obtain
\begin{equation}
    P_\text{loc}(\mathbf{x},\mathbf{k})=\int C\left(-\mathbf{k}+\frac{\mathbf{q}}{2},\mathbf{k}+\frac{\mathbf{q}}{2}\right)\,e^{i\mathbf{q}\cdot\mathbf{x}}\,\mathrm{d}^3q\;,
\end{equation}
and thus $P_\text{loc}(\mathbf{x},\mathbf{k})$ and
$C(\mathbf{k}_1,\mathbf{k}_2)$ are equivalent statistics (in the sense that they are related by a Fourier transform).
Finally, averaging over all possible positions $\mathbf{x}$ gives
\begin{align}
    \lim _{V\to \infty}\frac{1}{V}\,\int P_\text{loc}(\mathbf{x},\mathbf{k})\,\dif^3x&=\lim_{V\to \infty}\frac{(2\pi)^3\,C(\mathbf{k},-\mathbf{k})}{V}\nonumber \\&=
    \lim_{V\to \infty}\frac{\langle \left| \tilde{\delta}_\mc{V}(\mathbf{k})\right|^2\rangle}{ \,V}\;,
    \label{eq:avePrs}
\end{align}
which is the same expression we used to define the power spectrum in the translationally invariant case but is now evaluated using the galaxy overdensity in redshift space.

In order to compress the information on redshift-space distortions, we can expand the local power spectrum in Legendre polynomials while averaging over $\mathbf{x}$ 
\begin{equation}
    P_\ell(k)=\lim_{V\to \infty}\frac{2\ell+1}{V}\,\int P_\text{loc}(\mathbf{x},\mathbf{k})\,\mathcal{L}_\ell(\hat{\mathbf{k}}\cdot \hat{\mathbf{x}})\,
    \mathrm{d}^3x\;.
\end{equation}
Note that the monopole moment $P_0(k)$ coincides
with the average power spectrum we discussed in equation~(\ref{eq:avePrs}).

Natural estimators for the multipole moments of the (local) power spectrum
can be constructed by generalising the procedure we
introduced at the end of section~\ref{sec:Preal}. There is a drawback, however: namely that, due to the $\mathbf{x}$ dependence in the argument of the Legendre polynomials, these estimators cannot be generally expressed in terms of a product of Fourier transforms which makes them expensive to compute numerically \citep[see, however,][]{Yamamoto_estim, Scoccimarro2015, Bianchi2015, Hand+17}. The case $\ell=0$ is an exception to this rule as
it gives rise to the estimator

\begin{equation}
    \hat{P}_0(k) = 
    \frac{1}{V}\,\frac{\int_{V_k} \left| \tilde{\delta}_{\mc{V}}(\mathbf{k})\right|^2\,\dif^3k }{\int_{V_k}\mathrm{d}^3k}
    \;.
    \label{eq:P0est}
\end{equation}
Note that $\hat{P}_0(k)$ coincides with the standard estimator for the power spectrum used in the translation invariant case.

\subsection{Application to mock surveys}
\label{sec:mocksurveys}
In order to measure the monopole moment of the power spectrum from the mock
catalogues, we follow the same procedure which is used in
actual galaxy redshift surveys. Specifically,
we use the method introduced in \citet[][hereafter FKP]{FKP}
which provides an optimal weighting scheme
for estimating the power spectrum (under some assumptions). In this case, the unknown $\delta_\text{g}(\mathbf{x})$ is replaced with the effective field

\begin{equation}
\label{eq:Over_dens}
F(\mathbf{x})=\frac{I_{\mathcal S}(\mathbf{x})\,w(\mathbf{x})\,[n_\mathrm{g}(\mathbf{x})-\widehat{n}(\mathbf{x})]}
{\left[\int_{\mathcal S} w^2(\mathbf{x})\,\widehat{n}^2(\mathbf{x})\mathrm{d}^3x\right]^{1/2}}
\;,
\end{equation}
where, $I_{\mathcal S}$ is the indicator 
function of the region of space $\mc{S}$ (with comoving volume $S$) covered by the survey, 
$\widehat{n}(\mathbf{x})$ is an estimate of the galaxy density in the absence of clustering based on the survey data, and
$w(\mathbf{x})$ is a position-dependent statistical weight.
Note that the field $F$ can be written as $F(\mathbf{x})= W(\mathbf{x})\,\hat{\delta}(\mathbf{x})$
where $\hat{\delta}(\mathbf{x})=[n_\mathrm{g}(\mathbf{x})-\widehat{n}(\mathbf{x})]/\widehat{n}(\mathbf{x})$ is an estimate of
the local density contrast and
\begin{equation}
\label{eq:wind_func}
W(\mathbf{x})=\frac{I_{\mathcal{S}}(\mathbf{x})\,w(\mathbf{x})\,\widehat{n}(\mathbf{x})}{\left[\int_{\mathcal S} w^2(\mathbf{x})\,\widehat{n}^2(\mathbf{x})\,\mathrm{d}^3x\right]^{1/2}}
\end{equation}
defines the survey window function.
For each mock light cone,
we estimate $\widehat{n}(\mathbf{x})$ by measuring\footnote{In actual surveys, where the window function has a lot of small scale structure due to foreground objects and instrumental artefacts, the impact of $\widehat{n}(\mathbf{x})$ is often computed by generating a large number of `random' points that Poisson sample the selection function. We do not need this complication here.}
the mean density of galaxies within radial shells of width $\Delta x=20\,h^{-1}$ Mpc$^3$.
and by interpolating the results with a cubic spline.
Following FKP, we use
$w(\mathbf{x})=[1+\widehat{n}(\mathbf{x})\,\mc{P}]^{-1}$ inside the survey volume with $\mc{P}=20,000 \,h^{-3}$ Mpc$^{3}$. This scheme assumes that the galaxy distribution is a Poisson sampling of the underlying continuous density field $n_\text{g}(\bs{x})$. It provides equal weight per volume where we are limited by cosmic or sample variance ($\widehat{n}(\mathbf{x})\,\mc{P}\gg 1$) and equal weight per galaxy where we are limited by shot noise ($\widehat{n}(\mathbf{x})\,\mc{P}\ll 1$). This way it provides a
nearly minimum-variance estimate on the scales
where $P_0(k)\simeq \mc{P}$.

For the power-spectrum estimation via fast Fourier transform (FFT), we enclose the survey region ${\mathcal S}$ in a cubic box $\mc{V}$ with a comoving side of $L_{\mathrm{FFT}}=16\,h^{-1}$ Gpc
and sample the galaxy density at $1024^3$ points forming a regular Cartesian grid with linear spacing
$\Delta x=15.625\,h^{-1}$ Mpc.
Note that $L_{\mathrm{FFT}}$ is more than twice as large as the region covered by the survey at $z=1.8$.
This allows us to consider wavevectors with $k\geq k_{\mathrm{FFT}}=2\pi/L_\mathrm{FFT}\simeq 3.9\times 10^{-4}\,h$ Mpc$^{-1}$.
Although at first sight it might appear that $\Delta x$ is fairly large, taking also into account the function $\bar{n}_\text{g}(z)$ given in Fig.~\ref{Fig:functions} reveals that the mean number of galaxies found within one grid cell ranges between 4.23 and 0.63
when $0.9<z<1.8$. Therefore, many of the cells in the mock light cones are empty at high redshift.
We estimate 
the monopole moment of the galaxy power spectrum using
\begin{equation}
\hat{P}_0(k)=\frac{1}{L_\mathrm{FFT}^{3}}\,\frac{\int_{V_k} \left| \tilde{F}(\mathbf{k})\right|^2\,\dif^3k}{\int_{V_k}\mathrm{d}^3k}-P_\text{SN}\;,
\end{equation}
where the integrals are actually replaced by discrete sums over the individual FFT modes and
\begin{equation}
    P_\mathrm{SN}=\frac{\int \widehat{n}(x)\,w^2(\mathbf{x})\,\dif^3x}{\int \widehat{n}^2(x)\,w^2(\mathbf{x})\,\dif^3x}
\end{equation}
denotes the constant shot-noise contribution due to the fact that galaxies are discrete objects.
We compute $\hat{P}_0$ 
using bins of width $\Delta k=5\, k_\text{FFT}\simeq 1.96\times 10^{-3}\,h$ Mpc$^{-1}$
for all $5\times140$ mock catalogues
within four `tomographic' redshift bins with boundaries $\{0.9, 1.1, 1.3, 1.5, 1.8\}$ and also within the full redshift range.
In all cases, we consider both 
a Euclid-like survey and a full-sky survey. For the REAL and V$_\text{cmb}$ mocks, we also compute the spectra within a $30^{\circ}\times \,30^{\circ}$ `square' patch.

\subsection{Bias and scatter of $\hat{P}_0(k)$}
\label{sec:IC}
It is common knowledge that the FKP estimator suffers from several
drawbacks on the largest scales that can be probed in a survey. For instance, (i) it returns a smoothed version of the true power spectrum, (ii) it underestimates power due to the so-called integral constraint \citep{Peacock+Nicholson1991} and (iii) has correlated errors. 
In this section, we briefly discuss the impact of these effects
on our measurements.

A redshift survey probes the galaxy distribution within a finite (non-periodic) volume over which
the basis functions of the Fourier expansion (plane waves) are not orthonormal. 
Because of this,
the expectation value
of the FKP estimates, in the global plane-parallel
approximation,  
is
\begin{multline}
\label{eq:obs_power_mono}
\langle \hat{P}_0(k)\rangle=
\frac{\int_{V_{k}}
\left[\int P_0(q)\,|\widetilde{W}(\mathbf{k}-\mathbf{q})|^2\,\mathrm{d}^3q-P_\text{IC}(k)\right]  \dif^3 k}{\int_{V_{k}} \dif^3 k}
\end{multline}
where $P_0(k)$ denotes the `true' underlying monopole moment,
$\widetilde{W}(\mathbf{k})$ is the window function of the survey -- i.e. the Fourier transform of $W(\mathbf{x})$ -- and $P_\text{IC}(k)$ is a systematic error (known as the `integral constraint') that we discuss below. To begin with, let us focus on the first term within the square parentheses. In simple words,
the FKP estimator mixes different Fourier modes and computes a weighted average of the true power present in a realisation. 
Usually, the weight $|\widetilde{W}(\mathbf{k})|^2$ displays a series of peaks with
gradually decreasing amplitudes as $k$ grows.
The main peak is located around $k=0$ and has a width of $k_W \simeq S^{-1/3}$ (unless $\mc{S}$ is strongly elongated, in which case $k_W$ coincides with the reciprocal of the shortest dimension).

The systematic error $P_\text{IC}$ originates 
from the fact that, due to sample variance, the mean density of galaxies within the survey does not coincide with the actual cosmic density, i.e. $\widehat{n}\neq \bar{n}_\text{g}$.
Following a standard procedure, in section~\ref{sec:mocksurveys}, we have used the data themselves in order to 
estimate $\widehat{n}$, which is then inserted
in equation~(\ref{eq:Over_dens}) to get the effective density contrast.
This method
neglects the existence of large-wavelength fluctuations on scales comparable to and larger than the survey size and thus leads
to a biased estimate for $P_0(k)$.
The resulting integral-constraint bias is
\citep[e.g.][and references therein]{Peacock+Nicholson1991, deMattia:2019vdg}
\begin{equation}
    P_\text{IC}(k)=\frac{|\widetilde{W}(\mathbf{k}) |^2}{|\widetilde{W}(\mathbf{0})|^2} \int P_0(q)\,|\widetilde{W}(\mathbf{q})|^2\,\dif^3 q
\end{equation}
where the integral gives the underlying window-function-convolved monopole of the power spectrum evaluated at $k=0$.
\begin{figure}
    \centering
	\includegraphics[width=1\columnwidth]{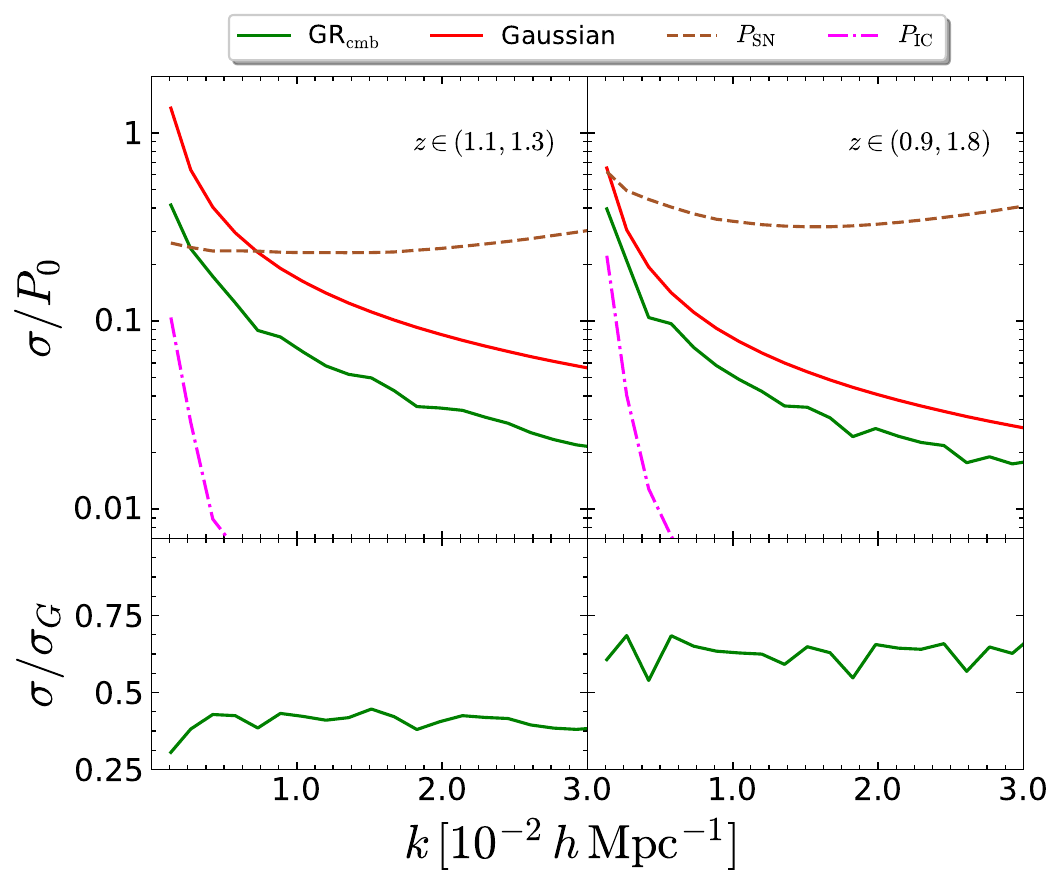}
    \caption{Top: The green solid line shows the
    relative rms scatter
    of the FKP estimator $\hat{P}_0(k)$ for a Euclid-like survey in the redshift bins $(1.1,1.3)$ (left) and $(0.9,1.8)$ (right). The curve is obtained by comparing the scatter and the mean of the estimates extracted from our 140 GR$_\text{cmb}$ mock catalogues.
    For comparison,
    the red solid line displays
    the noise level predicted using
    equation~(\ref{eq:sigma_FKP}) which assumes a Gaussian overdensity field and neglects
    the impact of the window function of the survey.
    The brown dashed line shows the systematic contribution due to shot noise, $P_\text{SN}$, which we subtract from our estimates. The magenta dash-dotted line refers to the systematic error generated by the integral constraint, $P_\text{IC}$.
    Bottom: Ratio between the rms scatter of the FKP estimator applied to the GR$_\text{cmb}$ mocks and the Gaussian approximation. }
    \label{fig:SCATTER_COMP}
\end{figure}
\begin{figure}
    \centering
	\includegraphics[width=1\columnwidth]{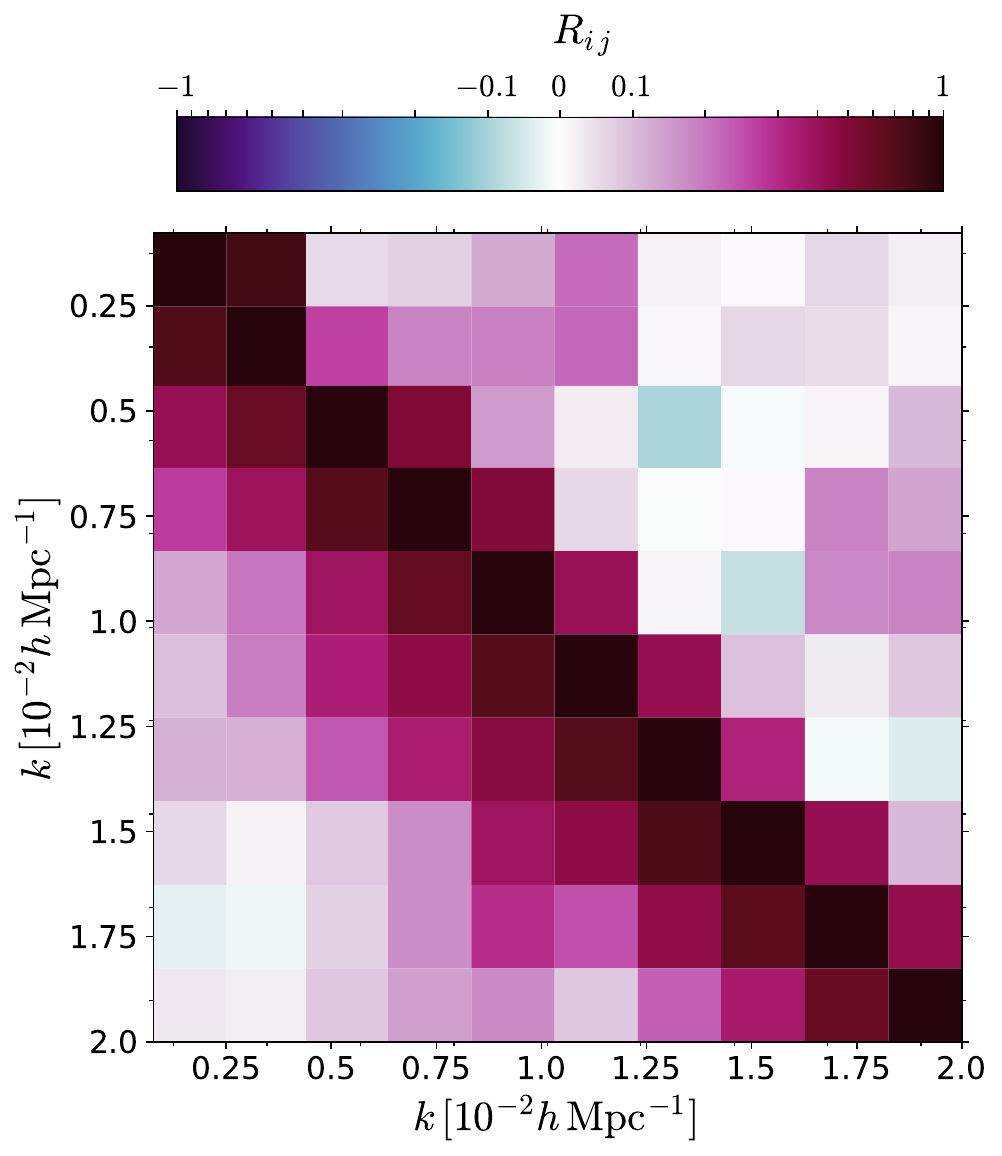}
    \caption{Correlation matrix $\mcor$ of the FKP estimates extracted
    from the GR$_\text{cmb}$ mocks in the redsfhit bin $(1.1,1.3)$ (lower triangular part) and $(0.9,1.8)$ (upper triangular part). Note that the latter has smaller off-diagonal elements due to the lower $k_W$.}
    \label{fig:CORR_MAT}
\end{figure}
Under the assumption of Gaussian density fluctuations and
for power bands of size $\Delta k\gg k_W$, the covariance matrix of FKP estimates is
diagonal with variance
\citep[e.g.][]{SCATTER_TEG}
\begin{equation}
\label{eq:sigma_FKP}
\frac{\sigma^2_\text{G}(k)}{P_0^2(k)} \simeq \frac{2\,(2\pi)^3}{V_{\mathrm{eff}}(k)\,V_k},
\end{equation}
where 
\begin{equation}
    V_{\text{eff}}(k)=\int_\mc{S} \left[\frac{\bar{n}(\bs{x})\,P_0(k)}{1+\bar{n}(\bs{x})\,P_0(k)}\right]^2\,\dif^3x
\end{equation}
denotes the effective volume probed (as the integrand is approximately one or zero 
 depending on whether the signal or shot noise dominate, i.e. $P_0\gg$ or $\ll \bar{n}^{-1}$) and $V_k$ is the volume of the $k$-shell centred around $k$ with width $\Delta k$ (if $k\gg \Delta k$, then $V_k \simeq 4 \pi k^2 \Delta k$). In general, further contributions to the covariance matrix arise from the connected trispectrum of the galaxy distribution convolved with the survey window function but these are subdominant on the very large scales we study here \citep[e.g.][]{Chan-Blot}. 
When $k$ and $\Delta k$ are comparable with $k_W$, however,
the action of the window function substantially reduces the variance of the FKP estimates with respect to the Gaussian approximation given in equation~(\ref{eq:sigma_FKP}) and increases the covariance among nearby $k$-bins. This is evident in  
Fig.~\ref{fig:SCATTER_COMP} which shows
clearly that the relative scatter $\sigma/\langle \hat{P}_0\rangle$ 
of the monopole estimates extracted from the Euclid-like
GR$_\text{cmb}$ mocks (green solid line)
is always smaller than expected from
equation~(\ref{eq:sigma_FKP}) ({red} solid line) on the large scales considered here. Comparing
the left and right panels provides evidence that
the ratio $\sigma/\sigma_\text{G}$
gets closer to one when $k_W$ is reduced
by considering broader redshift bins.
To further illustrate this point, in Fig.~\ref{fig:CORR_MAT} we show the correlation matrix of 
$\hat{P}_0(k)$ for the GR$_\text{cmb}$ mocks. Results for the redshift bins $(1.1,1.3)$ and $(0.9,1.8)$ are displayed in the lower and upper triangular parts, respectively. It is apparent that the amplitude of the off-diagonal elements grows as the window function gets broader.  

In Fig.~\ref{fig:SCATTER_COMP}, we also show $P_\text{SN}$ ({brown} dashed line). 
On the scales probed in this work, the systematic shift due to shot noise is always comparable to or larger than the statistical uncertainties but never dominates over the clustering signal.

In order to estimate
$P_\text{IC}(k)$ for a Euclid-like survey, we have re-computed all the power spectra for the GR$_\text{cmb}$ mocks by replacing $\widehat{n}$ in each light cone 
with its average over the 140 mocks. Our results for $P_\text{IC}(k)$ are shown with a magenta dash-dotted line in Fig.~\ref{fig:SCATTER_COMP}  and indicate that the bias is completely negligible with respect to the statistical uncertainties at  wavenumbers $k> \text{a few} \times 10^{-3} h$ Mpc$^{-1}$. In absolute terms, using $\widehat{n}$ leads to underestimating $P_0(k)$    by 5 to 20 per cent (depending on the redshift bin) on the largest scales 
we probe (where the random errors are at least a factor of a few larger). For this reason, in the remainder of this paper, we do not correct $\hat{P}_0(k)$ for the integral-constraint bias as this does not influence our conclusions.

\begin{figure*}
    \centering
	\includegraphics[width=1\textwidth]{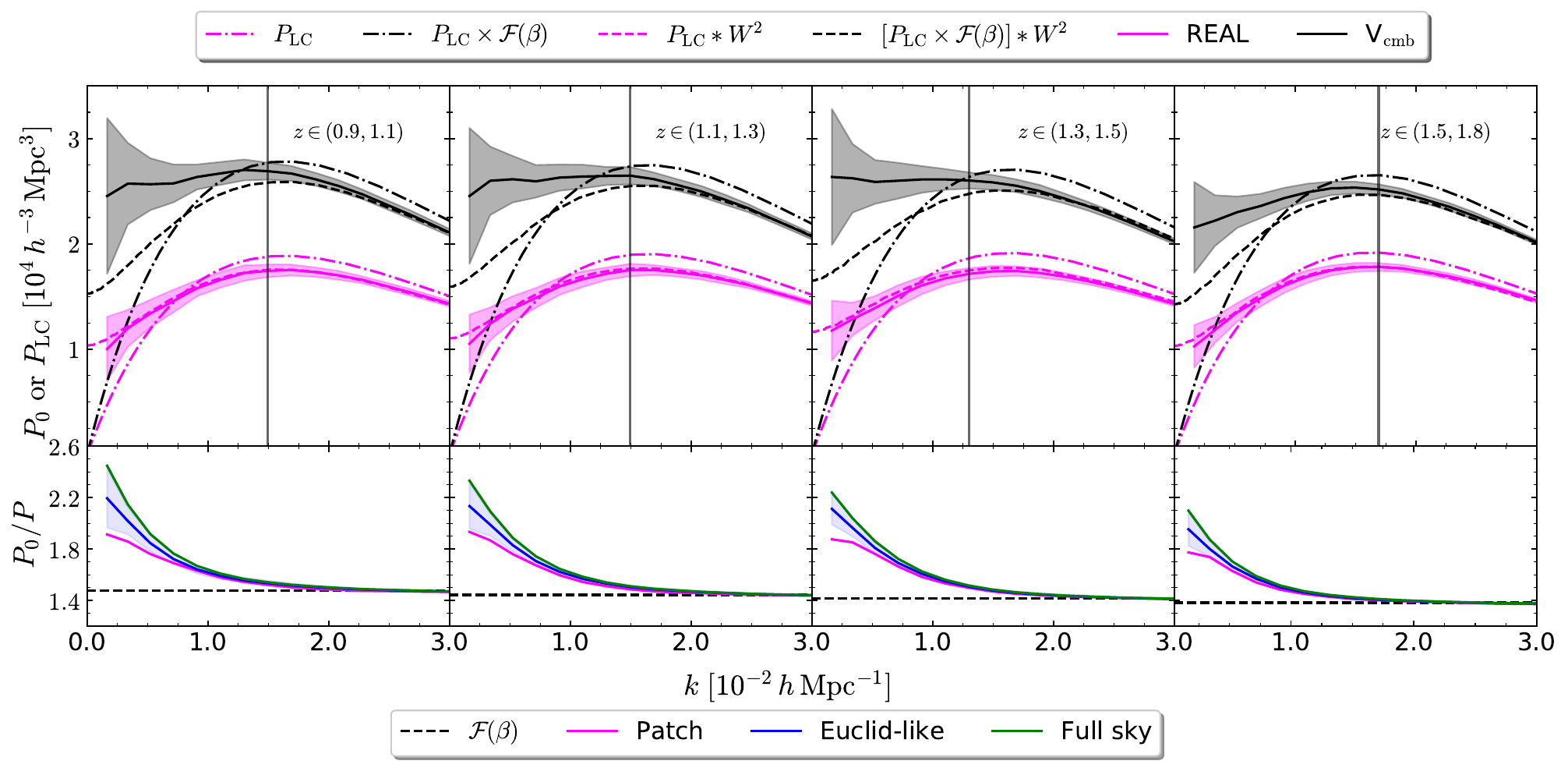}
	\caption{Top: Power spectra for a full-sky survey in different redshift bins (see Fig.~\ref{Fig:functions} for the properties of the galaxy population). The mean (solid) and the scatter (central 68-per-cent range, shaded region) of $P(k)$ obtained from the 140 REAL catalogs (magenta) and of $P_0(k)$ extracted from the V$_\text{cmb}$ mocks (black) are compared with the corresponding linear-theory predictions without (dash-dotted) and with (dashed) accounting for the survey window function. The vertical lines highlight the wavenumber at which the V$_\text{cmb}$ spectra deviate by more than one standard deviation from the Kaiser-boosted linear power spectrum. Bottom: The mean ratio $P_0(k)/P(k)$ between the monopole moment of the power spectrum in redshift space and the real space power spectrum for a full-sky survey, an Euclid-like survey, and a  $30^{\circ}\times \,30^{\circ}$ patch of the sky (from top to bottom) all with the same galaxy population. The horizontal dashed line indicates the Kaiser factor while the light shaded area denotes the central 68-per-cent range for the Euclid-like case.} 
    \label{fig:ratio}
\end{figure*}
\section{Results} \label{sec:results}
\subsection{Real space}
In the top panels of Fig.~\ref{fig:ratio}, 
we show the average (magenta solid line) and the scatter (central 68-per-cent range, pink shaded area) of the galaxy power spectrum $P(k)$ extracted from our 140 mock catalogues in real space. We consider a full-sky survey (for which we can derive an analytical window function) and the same redshift bins as in \citet{EuclidVII}. For comparison, we also plot 
the theoretical predictions obtained using standard 
perturbation theory at leading order ({magenta} dash-dotted line). This is obtained by (approximately) taking into account the time evolution within the light-cone as in \citet{Yamamoto_LC}, 
\begin{equation}
\label{eq:amp_power}
P_\text{LC}(k)=\left[\frac{\int_{z_{\rm{i}}}^{z_{\rm{f}}} b^2\, \bar{n}_{\rm{g}}^2\, D_+^2\,\frac{\dif V_\mc{S} }{\dif z}\,\dif z}{\int_{z_{\rm{i}}}^{z_{\rm{f}}} \bar{n}_{\rm{g}}^2\,\frac{\dif V_\mc{S}} {\dif z}\,\dif z}\right] \,P_\text{lin}(k)\;,
\end{equation} 
where $P_\text{lin}(k)$ denotes the linear matter power spectrum at $z=0$ obtained with the \ts{camb} code \citep{CAMBS}, 
$D_+(z)$ is the linear growth factor normalised
to one at the present time, and $V_\mc{S}$ is the comoving volume within the past light cone of the observer. Ultimately, we use equation~(\ref{eq:obs_power_mono}) -- actually its equivalent expression for $P(k)$ -- to
deal with the window function of the survey. The final result (black dashed line) is in extremely good agreement with the results obtained from the \ts{Liger} mocks. Note that the convolution with the window function flattens
the peak of the power spectrum by decreasing the clustering signal on scales $0.01\lesssim k \lesssim 0.04\, h$ Mpc$^{-1}$
and increasing it on the largest scales $k\lesssim 0.01\, h$ Mpc$^{-1}$.

\subsection{Wide-angle effects: accuracy of the Kaiser factor}

We build three-dimensional maps of the galaxy distribution by converting redshifts
into distances under the assumption of
a FLRW model universe with instantaneous expansion factor $a$ and Hubble parameter $H$.
This neglects the presence of perturbations
that influence the observed redshifts and thus distort the reconstructed maps. 
Considering only the radial contribution of peculiar velocities,
the observed redshift of a galaxy can
be expressed as
\begin{equation}
    1+z_\text{obs}=
    (1+z)\,(1+z_\text{pec})\,
    (1+z_\text{pec, obs})\;,
\end{equation}
where $z$, $z_\text{pec}$
and $z_\text{pec, obs}$
denote the cosmological redshift and the
corrections due to the peculiar velocity
of the galaxy and of the observer, respectively. For non-relativistic peculiar velocities $\bs{v}(\bs{x})$, this reduces to
\begin{equation}
    \frac{z_\text{obs}-z}{1+z}\simeq \frac{
    [\bs{v}(\bs{x})-\bs{v}_\text{obs}]\cdot \hat{\bs{x}}}{c}\;,
\end{equation}
which corresponds to a radial displacement
\begin{equation}
    \Delta\bs{x}\simeq
    \frac{[\bs{v}(\bs{x})-\bs{v}_\text{obs}]\cdot \hat{\bs{x}}}{a(z)H(z)}\,\hat{\bs{x}}=\left\{[\bs{u}(\bs{x})-\bs{u}_\text{obs}]\cdot \hat{\bs{x}}\right\}\,\hat{\bs{x}}\;,
    \label{eq:velshift}
\end{equation}
where
$\bs{u}(\bs{x})$ denotes
the peculiar velocity field expressed
in units of comoving length, i.e. divided by the factor $a(z)H(z)$.
The impact of this effect on galaxy clustering
was first quantitatively studied
in the pioneering
paper by \citet{Kaiser87} who showed that, in the continuum approximation and for linear perturbations, the galaxy overdensity
in redshift space can be written as 
\citep[see also][]{Hamilton-Culhane, Hamilton_review}
\begin{equation}
\delta_\text{obs}(\bs{x})=\delta(\bs{x})-
\frac{\partial [\bs{u}(\bs{x})\cdot \hat{\bs{x}}]}{\partial x}-\alpha(x)
    \,\frac{[\bs{u}(\bs{x})-\bs{u}_\text{obs}]\cdot \hat{\bs{x}}}{x} \;,
    \label{eq:kaiserfull}
\end{equation}
where $\delta(\bs{x})$ denotes the intrinsic overdensity (in real space), %
and\footnote{
With a little abuse of notation, we denote the composite function  $n\left(L_\textrm{min},z(x)\right)$ as $n\left(L_\textrm{min},x\right)$, and similarly for $L_\text{lim}$, $\mc{E}$, $H$ etc..}

\begin{align}\label{eq:alpha_c}
\alpha(x)&=2+\frac{\dif \ln \bar{n}_\text{g}(x)}{\dif \ln x}\nonumber \\
&=2+\left.\frac{\partial \ln n}{\partial\ln L_\textrm{min}}\right|_{L_\textrm{min}=L_\text{lim}(z)}\,\frac{\dif \ln L_\text{lim}}{\dif \ln x }+\frac{\partial \ln n}{\partial \ln x}\nonumber \\
&=2-Q \,\frac{\dif \ln L_\text{lim}}{\dif \ln x }+\frac{\partial \ln n}{\partial \ln (1+z)}\,\frac{\dif \ln (1+z)}{\dif \ln x}\nonumber \\
&=2-Q \,\frac{\dif \ln L_\text{lim}}{\dif \ln x }-\mc{E}\,\frac{H\,x}{c\,(1+z)}\nonumber \\
&=2\,(1-Q) -(2Q+\mc{E})\,\frac{H\,x}{c\,(1+z)}\nonumber\\
&\simeq 2(1-\mc{Q})\;,
\end{align}
where we have assumed a flat model universe (see appendix~\ref{sec:evmagbias}) and
the last equality holds true only in the limit $x\ll c/H$. 
The second term on the right-hand side of equation (\ref{eq:kaiserfull}) quantifies
how RSD squash or stretch galaxy separations due to the
velocity shear in the radial direction.
In the distant-observer approximation, 
this term boosts the power spectrum
in redshift space by the factor $(1+\beta \mu^2)^2$ where \begin{equation}
\beta=\frac{1}{b}\frac{\dif \ln D_+}{\dif \ln a}\,,
\end{equation}
and $\mu$ denotes the cosine of the angle between $\bs{k}$ and
the line of sight. For the monopole moment,
it gives an amplification factor of
\begin{equation}
    \mathcal{F}=1+\frac{2}{3}\beta+\frac{1}{5}\beta^2\;,
    \label{eq:kaiserboost}
\end{equation}
that from now on we will refer to as the Kaiser factor.
The last term on the right-hand side of equation (\ref{eq:kaiserfull}), instead,
gives the change in volume and in the 
mean density of galaxies induced by RSD.
This term
is usually neglected in theoretical models
as: (i) the shift generated by the rms peculiar velocity
is much smaller than the depth
of redshift surveys (a few Mpc vs hundreds or more), i.e.  $|\mathbf{u}-\mathbf{u}_\text{obs}|/x\ll 1$ \citep{Kaiser87,Hamilton-Culhane,1996ApJ...462...25Z} and (ii) the part of $\alpha$ that grows proportionally to $x$ matters only when $x$ is comparable to the Hubble radius (i.e. at high redshift). On such large scales, however, there are relativistic corrections that should be taken into account. We will revisit this issue later on in this paper.

\begin{figure*}
    \centering
	\includegraphics[width=1\textwidth]{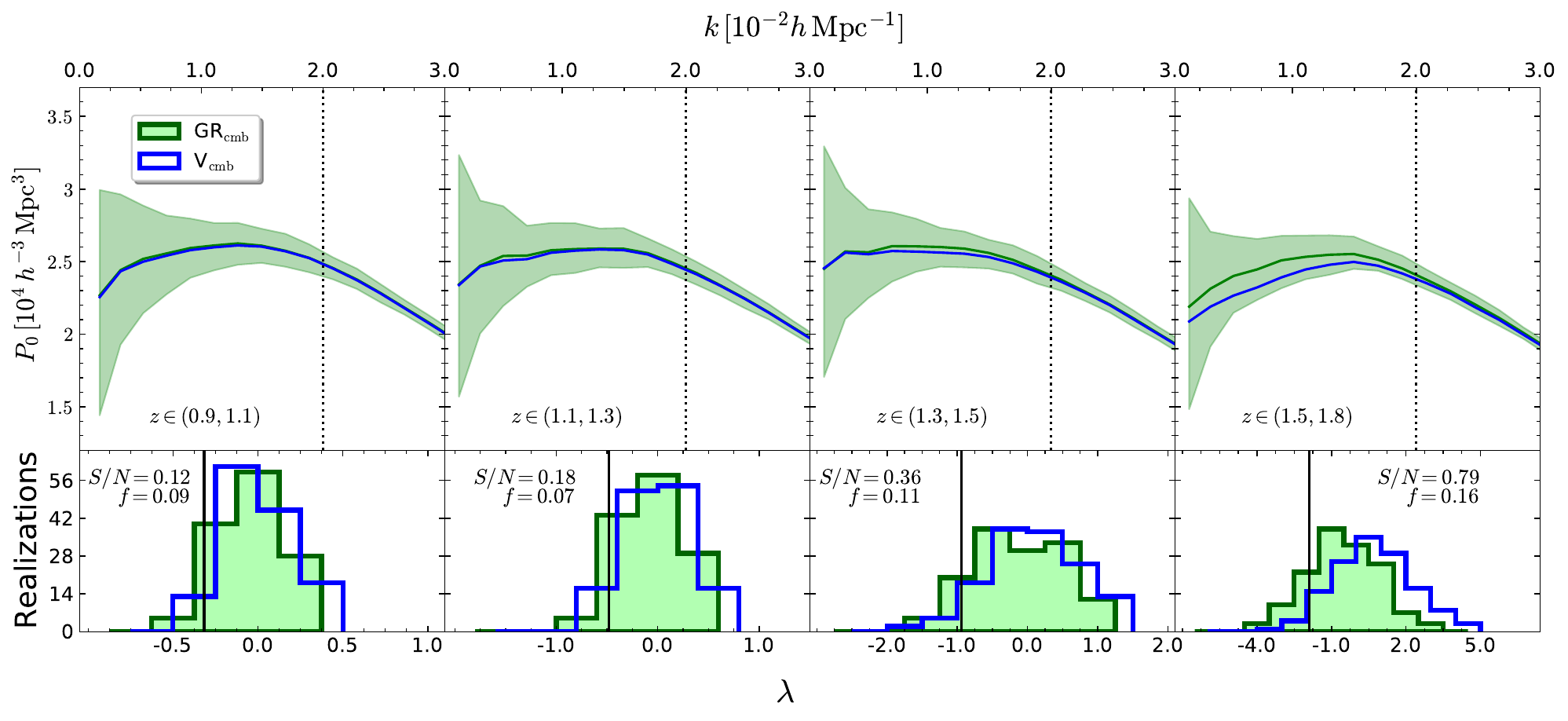}
    \caption{Top: Monopole moment of the power spectrum in different redshift bins for a Euclid-like survey.
    The mean signal
    obtained with the GR$_\text{cmb}$ mocks (green solid line)
    is compared with its counterpart computed from the V$_{\rm{cmb}}$ mocks (blue solid line). 
    The shaded region indicates the scatter among the GR$_\text{cmb}$ mocks (central 68 per cent region). The dotted vertical line marks the maximum $k$ used in the statistical test. 
    Bottom: Likelihood-ratio test between the models shown in the top panel (see the main text for details). The two histograms represent the PDF of the log-likelihood difference $\lambda$ for the GR$_\text{cmb}$ mocks (left, light green) and the V$_{\rm{cmb}}$ mocks (right, blue). The vertical line marks the decision threshold $\lambda_\text{c}$. Each panel reports the power of the test $f$ and the separation between the histograms in terms of the $S/N$ parameter.
    }
     \label{fig:CHI_WIDEANGLE}
\end{figure*}

The model delineated above (with $\alpha=0$)
has been routinely used to forecast the constraints that the Euclid mission will set on the cosmological parameters \citep[e.g.][]{Giannantonio+12,Majerotto+12,EuclidVII}.
Our study provides the opportunity to assess its accuracy
on large scales, where wide-angle effects should become important.
For this reason, in the top panels of
figure~\ref{fig:ratio}, we show the
monopole of the power spectrum
obtained considering the V$_\text{cmb}$ mock catalogues without shot noise (black solid line and grey shaded region). The clustering signal in redshift space is clearly enhanced with respect to real space. In order to facilitate comparison between the the simulations and equation~(\ref{eq:kaiserboost}),
we also plot
the Kaiser-boosted linear power spectrum for the light cone
before ({black} dash-dotted curve) and after ({black} dashed curve)
convolving it with the survey window function. 
Note that the window-convolved linear prediction agrees extremely
well with our simulations for $k\gtrsim 0.02\, h$ Mpc$^{-1}$
in all redshift bins.
On the other hand, the simulations show a clear increase of the clustering  amplitude on larger scales. The vertical line
in the plots highlights the wavenumber at which 
this systematic deviation exceeds the rms statistical error.
The bottom panels show the mean values over the 140 realisations of
the ratio $P_0(k)/P(k)$ as a function of $k$. The different colours
refer to a full-sky survey (green, top) an Euclid-like survey (blue, middle), and a $30^\circ \times 30^\circ$ patch of the sky (magenta, bottom). The light-blue shaded area indicates the uncertainty (central 68-per-cent range)
on the ratio based on the Euclid-like mocks and the horizontal dashed line corresponds to the Kaiser boost given in equation~(\ref{eq:kaiserboost}) (which does not depend on $k$).
Note that the discrepancy from the Kaiser formula becomes more and more severe as the survey size is increased as expected for wide-angle
effects.
In order to verify the origin of these corrections, we repeat the analysis using a special set of 
mock light cones that have
$\alpha=0$ by construction. Compared to Fig~\ref{fig:ratio}, we only note a tiny change in the lowest-$k$ bin while all the rest remains unchanged.
Based on this result, we conclude
that the excess power seen in Fig.~\ref{fig:ratio} is almost exclusively due to wide-angle effects.
For a Euclid-like survey
and wavenumbers $k\simeq \text{a few}\times 10^{-3} h$ Mpc$^{-1}$,
$P_0/P$ is approximately 47 per cent larger than the Kaiser factor and
scales as a power law $k^\gamma$
with $\gamma \approx-0.1$. It is worth mentioning that equation~\ref{eq:obs_power_mono} fully accounts for the window function of the survey in the global
plane-parallel appoximation while additional terms are required in
different approaches  \citep[e.g., in the local plane-parallel approximation,][]{Wilson+2017,Beutler++2017,Castorina:2017inr}

\begin{figure*}
    \centering
	\includegraphics[width=1\textwidth]{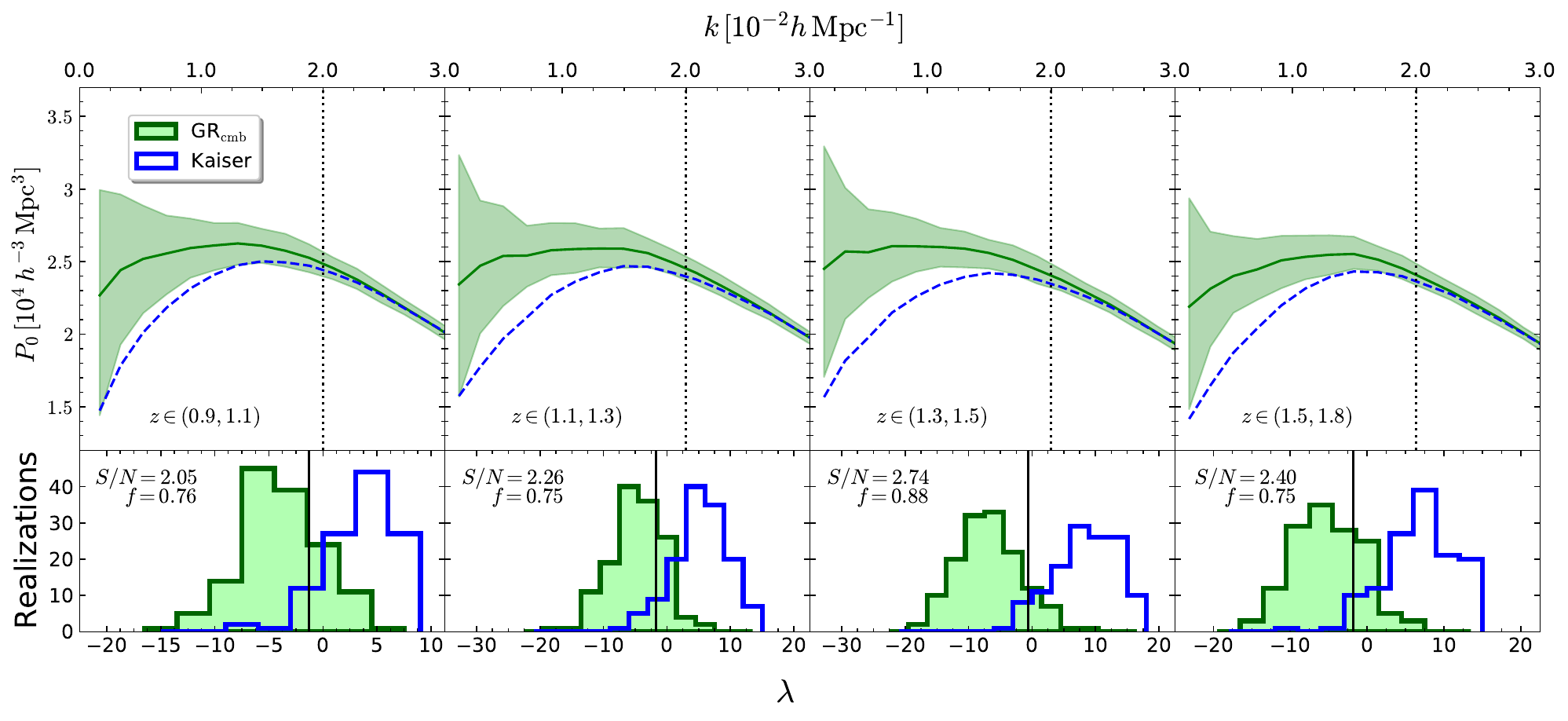}
    \caption{As in Fig.~\ref{fig:CHI_WIDEANGLE} but for the GR$_\text{cmb}$ mocks (green solid line) and the Kaiser model applied to the real-space data (blue dashed line). }
    
     \label{fig:CHI_KAISER}

\end{figure*}

\subsection{Gravitational lensing and other relativistic effects.}
Beyond peculiar velocities, several other effects generate RSD in a relativistic framework:
gravitional and Doppler lensing, standard and integrated Sachs-Wolfe effects (gravitational redshifts), Shapiro time-delay, etc.
\citep{Yoo:2009au, Bonvin-Durrer2011, Challinor:2011bk, Jeong:2011as}. 
In the top panel of Fig.~\ref{fig:CHI_WIDEANGLE}, we compare the power-spectrum monopole computed from the V$_\text{cmb}$ ({blue}) and GR$_\text{cmb}$ (green) mocks (now including shot noise). Shown are the mean spectra obtained averaging over the $140$ light cones. 
The shaded region indicates the scatter (central 68-per-cent range) for the GR$_\text{cmb}$ mocks. 
Since the corrections due to several relativistic effects (and, in particular, to lensing) scale proportionally to $Q-1$, we also compute the power spectrum for $Q=1$
(not shown in the figure as it basically coincides with that extracted from the V$_\text{cmb}$ mocks).
We notice that gravitational lensing slightly boosts the amplitude of the power-spectrum monopole on large scales but this systematic shift is always smaller than (at lower $z$) or comparable with (at higher $z$) the scatter between the mocks. 
The small difference between the power spectra extracted from the V$_\text{cmb}$ and GR$_\text{cmb}$ mocks follows from the
relatively flat luminosity function of the H$\alpha$ emitters detected at the Euclid flux limit which corresponds to values of the magnification bias of $2.0<Q(z)<2.6$ (see Fig.~\ref{Fig:functions}). 
Obviously, the discrepancy would be larger  if the faint-end slope of the luminosity function turned out to be substantially steeper than currently expected.

\subsection{Are these effects measurable?}\label{sec:STAT}

After having illustrated how different physical effects influence the monopole moment of the galaxy power spectrum, it is now time
to determine which of them are measurable and which are not.
We consider a Euclid-like survey and
use a `simple versus simple' hypothesis test (see also section 3.3.1 in paper I) as follows.
Let us denote by $\bs{d}$ a $n$-dimensional data vector containing
the monopole moment of the power spectrum extracted from one of our
GR$_\text{cmb}$ mock light cones. This represents one of the possible realisations
that can be measured by the Euclid-like survey.
We want to compare this particular measurement to different model predictions that correspond to the hypotheses we used to generate the 
V$_\text{cmb}$ and GR$_\text{cmb}$    mocks (all with no free parameters).
Assuming Gaussian errors, the likelihood 
that the data are generated by the model $M_i$ is
$\mc{L}(M_i|\bs{d})= \exp[-(\bs{d}-\boldsymbol{\mu}_i)^T \cdot \mcov_i^{-1} \cdot(\bs{d}-\boldsymbol{\mu}_i)/2]/[(2\pi)^{n/2} \det{\mcov_i}]$, where $\boldsymbol{\mu}_i$ and $\mcov_i$ denote the
expected data and the covariance matrix under the hypothesis that model $M_i$ is true. 
We estimate $\boldsymbol{\mu}_i$ and $\mcov_i$ by averaging over
the realisations of our mock light cones of the different types.
Eventually, we use the likelihood-ratio test statistic to compare two models
$\lambda = 2 \ln [\mc{L}(M_1|\bs{d})/\mc{L}(M_2|\bs{d})]=\chi^2_2-\chi^2_1+\ln \det \mcov_2-\ln \det \mcov_1$ with
$\chi^2_i=(\bs{d}-\boldsymbol{\mu}_i)^T \cdot \mcov_i^{-1} \cdot(\bs{d}-\boldsymbol{\mu}_i)$. If this quantity is positive, the data favour $M_1$ over $M_2$.
The logarithm of the likelihood ratio $\lambda$ is a random variable which assumes different values when the data set $\bs{d}$ is varied.
Its probability density function (PDF) plays a key role in inferential statistics. Obviously, the PDF depends on the data-generating process.
We consider two cases. First, we use the mock catalogs generated
under the hypothesis that model $M_1$ is true to obtain the
PDF $\mc{P}(\lambda|M_1)$. Then, we consider
the alternative hypothesis that $M_2$ is true and derive
$\mc{P}(\lambda|M_2)$.

In all our applications, we denote by $M_1$ the simpler model (which neglects some effects) and by $M_2$ the more complex
model (which includes all the effects we are considering).
We are now ready to set up a standard likelihood-ratio test
to test the simple null hypothesis that $M_1$ is true (against the simple alternative hypothesis that $M_2$ is).
Since $e^{\lambda/2}$ is small when $M_2$ is preferred by the data, we set a decision rule as: we reject the null hypothesis if $\lambda$ is
smaller than a predetermined threshold value, $\lambda_\text{c}$. We choose the latter
so that it corresponds to the significance level of 0.05 which gives the probability of making a type I error (i.e. of mistakenly rejecting the null hypothesis when it is actually true),
$\int^{\lambda_\text{c}}_{-\infty} \mc{P}(\lambda|M_1)\,\dif \lambda=0.05$.
The power of the binary test is given by the probability that the test
rejects the null hypothesis when the alternative one is true, i.e.
$f=\int^{\lambda_\text{c}}_{-\infty} \mc{P}(\lambda|M_2)\,\dif \lambda$ where $1-f$ gives the probability of making
a type II error (i.e. of wrongly failing to reject the null hypothesis). The closer $f$ is to one, the more powerful is
the test.
Basically, $f$ gives the fraction of the $M_2$ realisations in which
the test manages to reject $M_1$ at the 95 per cent confidence level.
Another piece of interesting information is how strong are typically the rejections of the null hypothesis. 
If $\mcov_1=\mcov_2$, one expects that
$\mc{P}(\lambda|M_2)=\mc{P}(-\lambda|M_1)$ where $\mc{P}(\lambda|M_i)$ is a Gaussian distribution with mean $m_i$ and rms  scatter $s_i=2 \sqrt{m_i}$. This leads to parameterize the separation between the histograms in terms
of the signal-to-noise ratio 
\begin{equation}
    S/N=(m_1-m_2)/s_1\,
    \label{eq:s/n}
\end{equation}
which reduces to
$\sqrt{m_1}$ in this particular case. Although, in general, the covariance matrices under the two hypotheses will not perfectly coincide (also because of the noise in their estimates) and the two PDFs will not be symmetric about $\lambda=0$, we  will keep using equation~(\ref{eq:s/n}) to quantify the characteristic significance level of the rejections. It is worth mentioning here that in all cases we find $s_i\simeq 1.9 \sqrt{m_i}$.

    

\begin{table*}
\label{tab:surveys}
	\begin{center}
	\caption{The $S/N$ and the power $f$ with which the Kaiser and V$_\text{cmb}$ models are ruled out by using the GR$_\text{cmb}$ mocks for a Euclid-like and a full-sky survey.
	These numbers quantify the need to account for wide-angle effects and weak-lensing magnification when modelling the data.
	The values listed in the left and right halves of the table are computed using the mock catalogues with and without shot noise, respectively. The right half, therefore, only accounts for the variations due to sampling different skies and gives an idea of the gain in $S/N$ from increasing the number density of galaxies with a measured redshift (e.g. by targeting fainter objects).} 
	\label{tab:S_N_TABLE}
  \begin{tabular}{ccccccccccccccccc}
  \hline
  &\multicolumn{8}{c}{Sampling errors and shot noise}&
  \multicolumn{8}{c}{Sampling errors only}\\
   $(z_\text{min},z_\text{max})$&\multicolumn{4}{c}{Euclid-like}&
  \multicolumn{4}{c}{Full sky}&
  \multicolumn{4}{c}{Euclid-like}&
  \multicolumn{4}{c}{Full sky}\\
  &\multicolumn{2}{c}{Kaiser} & \multicolumn{2}{c}{V$_\text{cmb}$}&\multicolumn{2}{c}{Kaiser} & \multicolumn{2}{c}{V$_\text{cmb}$}&\multicolumn{2}{c}{Kaiser} & \multicolumn{2}{c}{V$_\text{cmb}$}&\multicolumn{2}{c}{Kaiser} & \multicolumn{2}{c}{V$_\text{cmb}$}\\
            &$S/N$ & $f$  &$S/N$&$f$   &$S/N$ &$f$   &$S/N$ &$f$   &$S/N$&$f$  &$S/N$&$f$   &$S/N$&$f$  &$S/N$&$f$\\
       \hline
$(0.9,1.1)$ & 2.05 & 0.76 &0.12 &0.09  & 3.72 &0.99  & 0.15 &0.06  &2.21 &0.74 & 0.12&0.06  &4.11 &1.00 &0.17&0.06   \\ 
$(1.1,1.3)$ & 2.26 & 0.75 &0.18 &0.07  & 3.82 &0.99  & 0.22 &0.07  &2.31 &0.85 & 0.22&0.07  &4.24 &1.00 &0.34&0.09  \\ 
$(1.3,1.5)$ & 2.74 & 0.88 &0.36 &0.11  & 4.20 &1.00  & 0.59 &0.13  &3.15 &0.97 & 0.48&0.19  &4.55 &1.00 &0.70&0.16  \\ 
$(1.5,1.8)$ & 2.40 & 0.75 &0.79 &0.16  & 4.49 &1.00  & 1.48 &0.41  &2.93 &0.91 & 0.98&0.23  &5.54 &1.00 &1.84&0.62  \vspace{2mm}\\ 
$(0.9,1.8)$ & 1.91 & 0.56 &1.30 &0.41  & 3.48 &0.99  & 2.05 &0.69  &1.99 &0.74 & 1.47&0.41  &3.92 &1.00 &2.30&0.78 \\  
\hline
  \end{tabular}
  \end{center}
\end{table*}

An example of the application of this procedure is shown in Fig.~\ref{fig:CHI_KAISER}. In the top panel, the mean $P_0(k)$ 
extracted from the GR$_\text{cmb}$ mocks (green solid line) is compared with its counterpart obtained using the Kaiser boost  factor, 
\begin{align}
\overline{\mc{F}}=\frac{\int_{z_{\rm{i}}}^{z_{\rm{f}}} b^2\,\mc{F}\, \bar{n}_{\rm{g}}^2\, D_+^2\,\frac{\dif V_\mc{S} }{\dif z}\,\dif z}{\int_{z_{\rm{i}}}^{z_{\rm{f}}} b^2\, \bar{n}_{\rm{g}}^2\, D_+^2\,\frac{\dif V_\mc{S} }{\dif z}\,\dif z}\;,
\end{align}  
to enhance the real-space clustering measured from the simulations (blue dashed line). 
The shaded area indicates the scatter (central 68 per cent region) for the GR$_\text{cmb}$ mocks. The bottom panel shows the results for the likelihood-ratio test in which we compare the Kaiser ($M_1$) and the GR$_\text{cmb}$ ($M_2$) models.
We use $n=10$ bins\footnote{There are two reasons for limiting the number of bins to 10. First, all the effects we consider
in this work take place on very large scales. Second, our measurements of
the covariance matrices 
are based on the analysis of 140 mock catalogues
and it is well known that the estimates of the corresponding precision matrices (i.e. the inverse of the covariance matrices) get substantially biased if the number of bins is not much smaller than the number of realisations. 
We anyway correct for the mean bias in the estimated precision matrices by dividing them by the factor
$(N_\text{real}-1)/(N_\text{real}-n-2)\simeq 1.09$ \citep[][see also \citet{Hartlap+2007} for a first application in cosmology]{Kaufman67}.
} extending up to $k_\text{max}\simeq 0.02\,h$ Mpc$^{-1}$ (vertical dotted line).
The histogram on the right-hand side shows $\mc{P}(\lambda| M_1)$ (i.e. the PDF obtained using the Kaiser mocks) and the vertical line highlights $\lambda_\text{c}$. In this case, $\lambda$ is mostly positive
because model $M_1$ is preferred by these mock data. Note that, although
$M_1$ and $M_2$ systematically differ by an amount comparable to the
the statistical uncertainty, $|\lambda|$ is mostly smaller than $n$
because of the strong correlation between
the data points at different wavenumbers induced by the window function of the survey.
The histogram on the left-hand side, instead, represents
$\mc{P}(\lambda| M_2)$ which is  shifted towards negative values of $\lambda$ as the GR$_\text{cmb}$ model tends to be preferred by the GR$_\text{cmb}$ mocks.
Since the two histograms partially
overlap,
the likelihood-ratio test
applied to the GR$_\text{cmb}$ mocks (which mimic actual observations)
is able to reject the Kaiser model  (i.e. gives values $\lambda<\lambda_\text{c}$) 
in approximately three quarters of the realisations. The power of the test is reported in each panel together with the $S/N$.
Based on these results, we conclude that, for $k<0.02\, h$ Mpc$^{-1}$, the Kaiser model for RSD does not provide an accurate description of $P_0(k)$ in a Euclid-like survey and can be statistically ruled out in about 75 to 88 per cent of our mock samples.
The situation becomes even more extreme by considering a full-sky survey. In this case (not shown in the figure), we find that $S/N$ grows in all redshift bins and assumes values up to $4.49$ corresponding to $f=1.00$ (see the left half of Table~\ref{tab:S_N_TABLE}).
In order to have a rough idea of future perspectives achievable with `denser' surveys, we also
report results obtained neglecting shot noise (right half of Table~\ref{tab:S_N_TABLE}). These entries clearly illustrate that the Kaiser model is not suited to describe the large-scale power-spectrum monopole 
obtained by targeting fainter galaxies.

\begin{figure}
    \centering
	\includegraphics[width=0.4\textwidth]{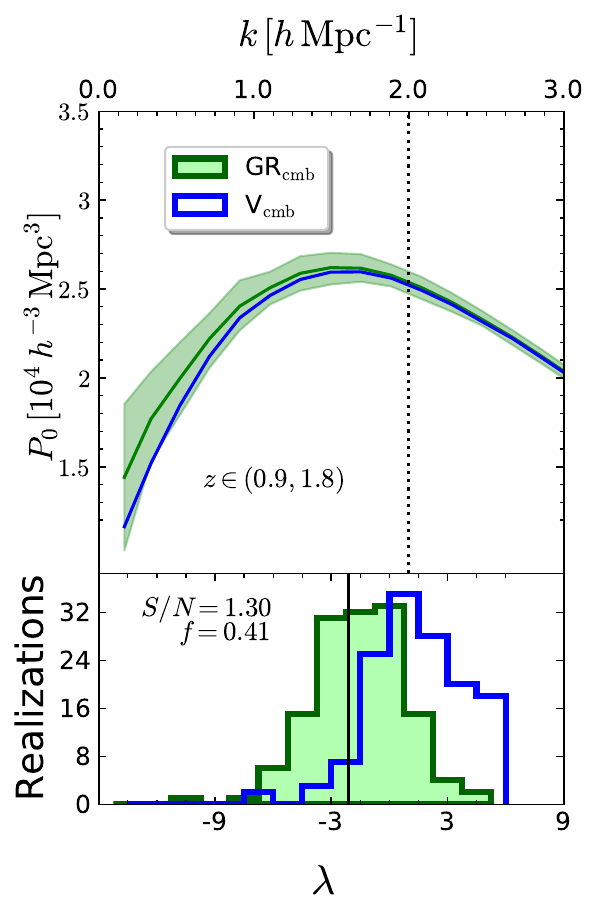}
    \caption{As in Fig.~\ref{fig:CHI_WIDEANGLE}  but for a broader redshift bin.}
\label{fig:CHI_BIGBIN}
\end{figure}

In Fig.~\ref{fig:CHI_WIDEANGLE} (see also Table~\ref{tab:S_N_TABLE}), we repeat the analysis by considering the GR$_\text{cmb}$ and V$_\text{cmb}$ mocks.  It turns out that, in most cases, we cannot discriminate between the models based on the data of a Euclid-like survey. The power of the likelihood-ratio test is rather low and grows a bit only in the highest-redshift bin.
All this suggests that
accounting for radial redshift-space distortions due to peculiar velocities (i.e. dropping the distant-observer approximation in the Kaiser model) should be sufficiently accurate
to describe $P_0(k)$ at least for $z<1.5$. At higher redshifts and for $k<0.02\,h$ Mpc$^{-1}$, however, this model would be systematically shifted by an amount
that roughly corresponds to the 
{sampling uncertainty of $P_0(k)$ (neglecting shot noise).} This bias in the simplified models might hinder cosmological inference based on the clustering signal at very large scales. The importance
of this shift due to relativistic effects should be evaluated case
by case.
This result reflects the low magnification bias of the Euclid galaxies and might change if the faint-end slope of their luminosity function turns out to be steeper than currently estimated.
\begin{figure*}
    \centering
	\includegraphics[width=1\textwidth]{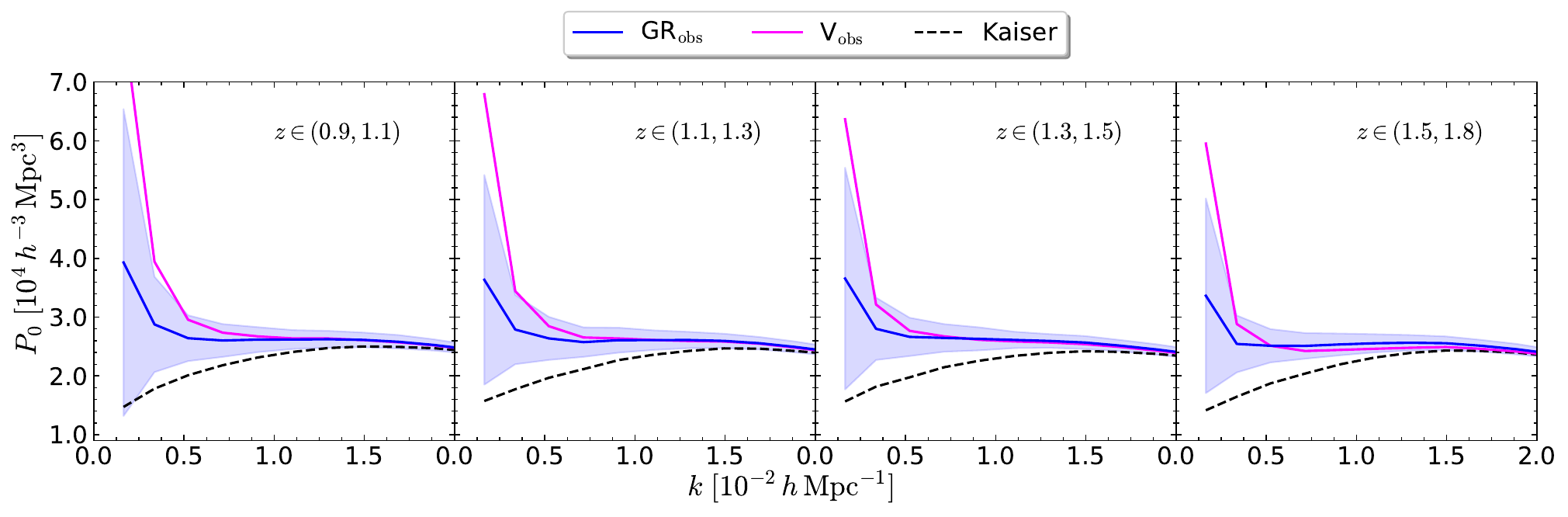}
\caption{As in the top panels of Fig.~\ref{fig:CHI_WIDEANGLE} but in the observer's rest frame. The solid curves show the mean spectra extracted from the  GR$_\text{obs}$ (blue) and V$_\text{obs}$ (magenta) mocks. The shaded regions indicate the central
68-per-cent scatter for the GR$_\text{obs}$ case.
The dashed curves display the spectra
obtained using the Kaiser model (see the blue dashed curves in \ref{fig:CHI_KAISER}). 
Note that the this figure and the top panels in Fig.~\ref{fig:CHI_WIDEANGLE} show different ranges in $P_0$ in order to improve readability.
} 
 \label{fig:all_effects_euc}
\end{figure*}
The conclusions we have drawn above apply to the relatively narrow redshift bins of width $\Delta z \simeq$ 0.2-0.3. 
In Fig.~\ref{fig:CHI_BIGBIN}  (see also the last row in Table~\ref{tab:S_N_TABLE}), we repeat the analysis for 
the entire redshift interval covered by the Euclid spectroscopic survey: $0.9<z<1.8$. In this case,
we find that the impact of the light-cone effects (and, in particular, of gravitational lensing) is detectable in 41 per cent of the realizations. This result shows that the importance of lensing effects increases with the size of the redshift bin under consideration.
On the other hand, Table~\ref{tab:S_N_TABLE} shows that wide-angle effects are detected with a lower $S/N$ in the broader redshift bin.
This likely reflects the larger contribution to the clustering signal from galaxy
pairs with (mostly) radial separations with respect to narrower redshift bins (in which the signal at small $k$ is generated by pairs with large angular separations). Indeed, we find that considering broader and broader redshift bins with the same mean redshift results into a reduced clustering signal on large scales.

In summary, the main conclusion we draw from the data listed in Table~\ref{tab:S_N_TABLE} is threefold.
First, the $S/N$  generally increases with the size of survey. 
  Second,  the lensing magnification has a discernible effect on $P_0(k)$ if broad redshift bins of width $\Delta z\simeq 1$ are considered particularly in a full-sky survey. 
  Third, the contribution of the shot noise to the error budget is small compared to that arising from the sampling errors. This is due to the high number density of galaxies measured by a Euclid-like survey.

A note is in order here.
It is plausible that the statistical significance of the detection
of the different effects is influenced by the choice of
the power-spectrum estimator.
Based on the plans of the
Euclid collaboration, our study utilises the FKP method. On the large scales probed in this work, however, the FKP estimates and their covariance matrix are heavily affected by
the window function of the survey.
For this reason, it is reasonable to expect improvements in the $S/N$ when
window-less quadratic estimators \citep[e.g.][]{SCATTER_TEG, Philcox21}
and/or weighting schemes which depend on the properties of the galaxy population \citep[e.g.][]{Smith+Marian15, Castorina+19} are employed.

\subsection{The finger of the observer effect}
\label{sec:vobs}

The power-spectrum monopole extracted from the V$_\text{obs}$ ({magenta}) and GR$_\text{obs}$ (blue) mocks are shown 
in Fig.~\ref{fig:all_effects_euc}. 
Three things are worth noticing from the comparison with the top panels in Figs.\ref{fig:CHI_KAISER} and 
\ref{fig:CHI_WIDEANGLE}.
First, the monopole of the power spectrum measured in the standard of rest of the observer (hereafter, observer frame)
obviously differs from its analogue computed in the standard of rest in which the CMB is isotropic (CMB frame). In particular, the spectra in the observer frame show a substantially higher clustering amplitude on the largest scales where $\dif \log P_0(k)/\dif \log k$ ranges from $-0.39$ to $-0.44$, depending on the redshift bin. This signal could be confused with the signature of primordial non-Gaussianity
convolved with the window function.
Second, the difference between the
spectra obtained by accounting for all the relativistic effects and by only considering the peculiar velocities is larger in the observer frame.
Third, the scatter of $P_0$ is also larger in the observer frame.
In the remainder of this section, we clarify the origin of these differences.

\begin{figure}
    \centering
    \hspace*{-1cm} \includegraphics[width=1.1\columnwidth, keepaspectratio=true]{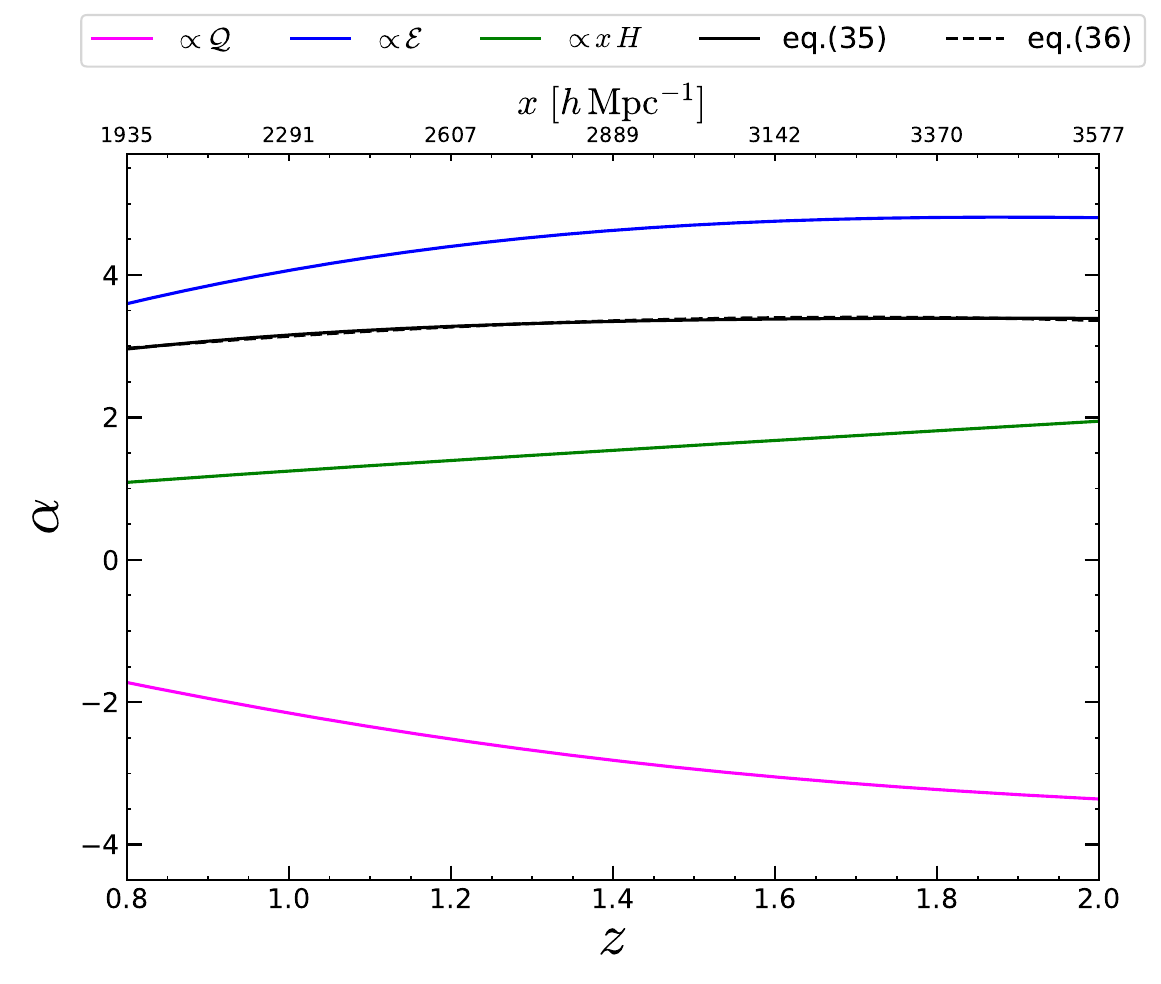}
    \caption{{The $\alpha$ function  (black solid) and its different components plotted with a colour scheme matching equation~(\ref{eq:alphaGRobs}). The dashed black line (almost perfectly overlapping with the solid one) displays the quadratic fit to $\alpha$ given in equation~(\ref{alpha_fit}).   }}
    \label{fig:ALPHAS}
\end{figure}

In a perturbed FLRW model,
the observed redshift of a source
is influenced by the peculiar velocity
and potential of the observer that thus generate RSD. Neglecting
the integral contributions (see equation~(8) in paper I for the full expression), this
corresponds to a radial shift of
\begin{equation}
    \Delta \bs{x}\simeq\frac{1}{a(z)H(z)}\, \left[\frac{\phi_\text{obs}}{c}-\bs{v}_\text{obs}\cdot\hat{\bs{x}}\right]\,\hat{\bs{x}}\;,
    \label{eq:kaiserfullvobs}
\end{equation}
where $c$ denotes the speed of light.
On linear scales,
the velocity contribution is much more important than that from the potential
which justifies the use of equation~(\ref{eq:velshift}).
From equation~(\ref{eq:kaiserfull})
we deduce that the galaxy overdensities
in the CMB and observer frames differ
by a distance-dependent dipole term \citep{Hamilton-Culhane, Hamilton_review}
\begin{align}
    \delta_\text{obs}(\bs{x})&=
    \delta_\text{cmb}(\bs{x})+\delta_\text{dip}(\bs{x})=
    \delta_\text{cmb}(\bs{x})+\frac{\alpha(x)}{x}\,\bs{u}_\text{obs}\cdot \hat{\bs{x}}\nonumber \\
    &=\delta_\text{cmb}(\bs{x})+\frac{\alpha(x)}{x}\,\frac{\bs{v}_\text{obs}\cdot \hat{\bs{x}}}{a[z(x)]\,H[z(x)]}
    \;.
    \label{eq:dipoleobs}
\end{align}
Accounting for the linear relativistic corrections leaves this
equation unchanged provided one replaces equation~(\ref{eq:alpha_c}) with\footnote{It is worth mentioning here that, contrary to equation~(\ref{eq:kaiserfull}), in the relativistic case, the overdensities generated by $\bs{v}(\bs{x})$ 
scale with a different $\alpha$ function
\citep[e.g.][]{Challinor:2011bk, Bertacca:2012tp, Raccanelli:2013dza, Raccanelli:2016avd} 
\begin{align}
     \alpha(x)&= 
      2(1-\mc{Q})  +\frac{x \, H}{c\, (1+z)}\left[ -\mc{E}+1+2\mc{Q} - \frac{1+ z}{H}\, \frac{\dif H}{\dif z}\right]\,.\nonumber
\end{align}}
\citep{Maartens+2018, Bertacca:2019wyg}
\begin{align}\label{eq:alphaGRobs}
     \alpha(x)&=   \textcolor{magenta}{2(1-\mc{Q})}  \textcolor{blue}{\,-\,\frac{x \, H}{c\, (1+z)}\, \mc{E}} \textcolor{ForestGreen}{\,+\,\frac{x \, H}{c\, (1+z)}\left[ 3 - \frac{1+ z}{H}\, \frac{\dif H}{\dif z}\right]}\,,
\end{align}
where the dependence of $z$ on $x$ is acknowledged. In Fig.~\ref{fig:ALPHAS} we show the redshift evolution of the different terms for a Euclid-like survey using the same
colours as in the equation above.
A partial cancellation takes place between the negative magenta term and the positive green term so that, in this case, the predominant contribution to $\alpha$ comes from the evolution bias. 
\begin{figure*}
    \centering
	\includegraphics[width=1\textwidth]{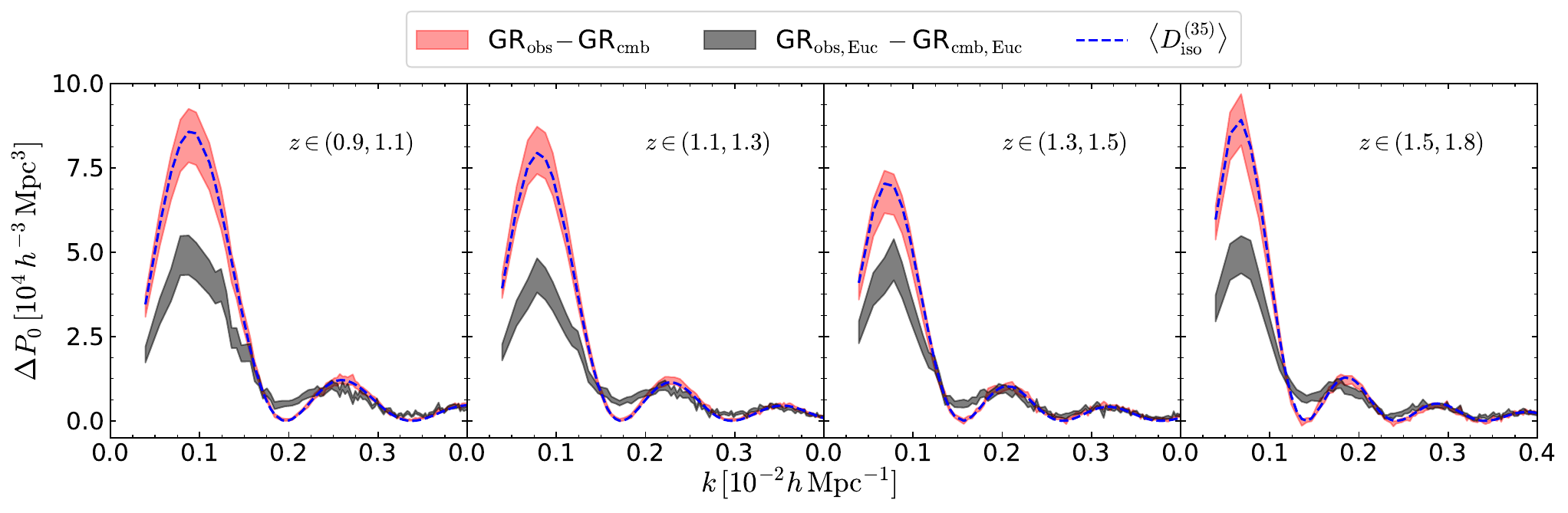}
    \caption{The difference between the monopole moments of the power spectra measured in the observer's rest frame and in the CMB frame -- see equation~(\ref{eq:deltaP0def}) --
    in different redshift
    bins. The shaded regions indicate the uncertainty range for the mean computed from different sets of mock catalogues -- namely, considering the difference between the GR$_\text{obs}$ and GR$_\text{cmb}$ mocks in a full-sky survey (salmon pink). The black shading and in a survey with a Euclid-like footprint (grey).
    The blue dashed line ,instead, refers to equation~(\ref{eq:dipole_boost}) 
    with the $\alpha$ function given in equation (\ref{eq:alphaGRobs}) respectively.
    In all cases, we use the selection function for a Euclid-like survey.
    }
    \label{fig:j12}
\end{figure*}

Fig.~\ref{fig:ALPHAS} also nicely illustrates the importance of accounting for the green term when dealing with high-redshift samples.
For $0.8<z<2$,
the resulting $\alpha$ function (solid black line) is well approximated by the quadratic relation
(dashed line)
\begin{equation}
    \alpha \simeq -0.53\, z^2 + 1.81\, z + 1.86\;.
    \label{alpha_fit}
\end{equation}

By analogy with the `fingers of God' effect \citep{Jackson72},
we refer to the dipole 
imprinted on to the galaxy distribution 
in redshift space
as the `finger of the observer' effect.
In fact, $\delta_\text{dip}$ points towards the direction of motion of the observer.
There is some confusion in the literature about the name of this effect.
In the 1990s, several studies investigated whether the peculiar velocity of the Local Group of galaxies inferred from the CMB dipole could be explained in terms of
the linear gravitational acceleration 
generated by the observed galaxy distribution in a flux-limited survey.
Since the conversion of galaxy redshifts into distances depends on the peculiar velocity of the Local Group,
it turns out that errors in $\mathbf{v}_\text{LG}$ 
generate a bias in the estimated acceleration vector \citep{Kaiser87, Kaiser+Lahav}.
The names `Kaiser effect' \citep{Strauss+92} and `rocket effect' \citep[e.g.][]{Nusser+Davis, Hamilton_review, Schmoldt+99a} have been used to indicate   
the spurious acceleration measured by a non-geodesic observer (e.g. an observer in a rocket). \citet{Bertacca:2019wyg} used the same term to refer to $\delta_\text{dip}$. Although the two effects are related, here, we prefer to distinguish between the bias in the reconstructed acceleration due to errors in $\mathbf{v}_\text{LG}$ (the Kaiser or rocket effect) and the actual kinematic dipole overdensity measured in the observer's frame (the finger of the observer effect).

We want to compute the contribution
to the power spectrum generated by cross correlating $\delta_\text{dip}$ with itself for a given observer and a full-sky survey.
It follows from the definitions above that this quantity can be written as
\begin{equation}
    D(\mathbf{k})  = \frac{\int A_1\,A_2\,  (\bs{v}_\text{obs}\cdot \hat{\bs{x}}_1)\,(\bs{v}_\text{obs}\cdot \hat{\bs{x}}_2) \,e^{i\bs{k} \cdot (\bs{x}_2-\bs{x}_1) }\,\dif^3 x_1 \,\dif^3 x_2}{\int \bar{n}^2(x)\,\dif^3 x}
    \label{eq:deltap0}
\end{equation}
with $A_i=\bar{n}_i\,\alpha_i/(x_i\,a_i\,H_i)$. Note that no ensemble average needs to be taken here as there are no stochastic terms appearing in the equation.
Let us first concentrate on the integrals
of the type $\int_{4\pi}(\hat{\mathbf{v}}_\text{obs}\cdot \hat{\mathbf{x}})\,e^{\pm i\bs{k} \cdot \bs{x}}\,\dif^2 \Omega_{x}$
over the solid angles spanned by $\mathbf{x}$. Using the plane wave expansion 
\begin{equation}
    e^{i\bs{k}\cdot\bs{x}} = \sum_\ell(2\ell+1)\, i^\ell\, j_\ell(kx)\, \mc{L}_\ell(\hat{\bs{k}}\cdot\hat{\bs{x}})
\end{equation}
(where $j_\ell$ and $\mc{L}_\ell$ denote the spherical Bessel functions and the Legendre polynomials, respectively),
together with the identities 
$\hat{\bs{q}}\cdot \hat{\bs{x}}_1=\mc{L}_1(\hat{\bs{q}}\cdot \hat{\bs{x}}_1)$ and
\begin{equation}
     \int \mc{L}_\ell(\hat{\bs{q}}\cdot \hat{\bs{x}}) \mc{L}_m(\hat{\bs{k}}\cdot \hat{\bs{x}})\,\dif^2 \Omega_{x} = \frac{4\pi}{2\ell+1}\,\delta^\text{K}_{\ell m}\, \mc{L}_m(\hat{\bs{q}}\cdot \hat{\bs{k}})
 \end{equation}
 (where $\delta^\text{K}_{lm}$ denotes the Kronecker delta),
we obtain
\begin{align}
&\int_{4\pi} 
(\hat{\mathbf{v}}_\text{obs}\cdot \hat{\mathbf{x}})\,e^{\pm i\bs{k} \cdot \bs{x}}\,\dif^2 \Omega_{x} \nonumber \\
&=
\sum_\ell(2\ell+1)\, (\pm i)^\ell\, j_\ell(kx)\, \int_{4\pi} (\hat{\mathbf{v}}_\text{obs}\cdot \hat{\mathbf{x}})\, \mc{L}_\ell(\hat{\mathbf{k}}\cdot\hat{\mathbf{x}})\,
\dif^2 \Omega_{x}
\nonumber \\
&=\pm 4\pi\,i\, j_1(kx)\, (\hat{\mathbf{v}}_\text{obs}\cdot \hat{\mathbf{k}})\;.
\end{align}

Therefore, the `power spectrum' of the dipole term in a single realisation
can be expressed as
\begin{align}
D(k, \hat{\mathbf{v}}_\text{obs}\cdot \hat{\mathbf{k}})  = 16 \pi^2\,\frac{(\mathbf{v}_\text{obs}\cdot \hat{\mathbf{k}})^2}{H_0^2}\,
\frac{I^2(k)}{\int \bar{n}^2(x)\,\dif^3 x}\;,
\label{eq:dipolespectrum}
\end{align}
where 
\begin{equation}
I(k)=\int  \frac{x\,\bar{n}\,\alpha}
{a\, E }\,j_1(k x)\,\dif x\;,  
\label{eq:Idef}
\end{equation}
with $E=H/H_0$.
For a narrow shell at distance $x$, we get
\begin{equation}
   D_\text{thin}(k, \hat{\mathbf{v}}_\text{obs}\cdot \hat{\mathbf{k}})\simeq 4\pi\,\frac{(\mathbf{v}_\text{obs}\cdot \hat{\mathbf{k}})^2}{a^2 H^2}\,\alpha^2\,j_1^2(kx)\;.
\end{equation}
Obviously, $D$ is anisotropic as it depends both on $k$ and the square of the cosine of the angle between the wavevector and the peculiar velocity of the observer.
Averaging over all the directions of the wavevectors in order to get an isotropic spectrum gives
\begin{align}
D_\text{iso}(k)  = \frac{16 \pi^2}{3}\,\frac{v_\text{obs}^2}{H_0^2}\,
\frac{I^2(k)}{\int \bar{n}^2(x)\,\dif^3 x}\;,
\label{eq:Diso}
\end{align}
while further averaging $v^2_\text{obs}$ over an ensemble of observers gives 
\begin{align}
\langle D_\text{iso}(k) \rangle = 16 \pi^2\,\frac{\sigma^2_v}{H_0^2}\,
\frac{I^2(k)}{\int \bar{n}^2(x)\,\dif^3 x}\;,
\label{eq:dipole_boost}
\end{align}
where $\sigma_v$ denotes the one-dimensional peculiar velocity dispersion 
of the observers. Similarly,
\begin{align}
\langle D(k, \hat{\mathbf{v}}_\text{obs}\cdot \hat{\mathbf{k}})\rangle  = 48 \pi^2\,\sigma^2_v\,\frac{(\hat{\mathbf{v}}_\text{obs}\cdot \hat{\mathbf{k}})^2}{H_0^2}\,
\frac{I^2(k)}{\int \bar{n}^2(x)\,\dif^3 x}\;,
\label{eq:meandipolespectrum}
\end{align}
where we have assumed that the
three-dimensional velocity dispersion
is a factor $\sqrt{3}$ larger than its one-dimensional counterpart (i.e. that $\hat{\mathbf{v}}_\text{obs}$ is isotropically distributed).

Throughout this section, we work under the hypothesis that $D$ dominates over the term generated by the cross correlation between $\delta_\text{dip}$ and $\delta_\text{cmb}$. Therefore,
the difference between the ensemble-averaged $P_0$ measured in the observer frame and in the CMB frame is
\begin{align}
   \Delta P_0(k)=P_0^{(\text{obs})}(k)-P_0^{(\text{cmb})}(k)\simeq D_\text{iso}(k)\;.
   \label{eq:deltaP0def}
   \end{align}
In Fig. ~\ref{fig:j12}, we verify that this expression accurately describes the measurements from the mock catalogues obtained with \ts{LIGER}. To do this, we re-compute the power spectra with the finest possible resolution in $k$ (i.e. by just averaging over the directions for all the individual values of $k$ that appear in the FFT grid). 
The standard uncertainty of the 
mean $\Delta P_0$ over the $140$ mocks is shown with a salmon-pink shaded region for a full-sky survey and with a grey one for a Euclid-like survey. Note that the high sampling in $k$-space allows us to clearly detect the damped oscillatory features that are not visible in Fig.~\ref{fig:all_effects_euc}
due to the broader bins with 
$\Delta k= 5\, k_\text{FFT}$
(in this case, the left-most bin extends beyond
the first zero of $\Delta P_0$).
The blue dashed line is obtained using equation (\ref{eq:dipole_boost}) with the actual velocity dispersion of the observers in the mock catalogues.

The analytical model accurately matches the outcome of the simulations for the full-sky survey while the results obtained with the Euclid footprint are obviously affected by the widow function which suppresses the amplitude of the first peak and alters the shape around the first minimum.

\begin{figure*}
    \centering
	\includegraphics[width=1\textwidth]{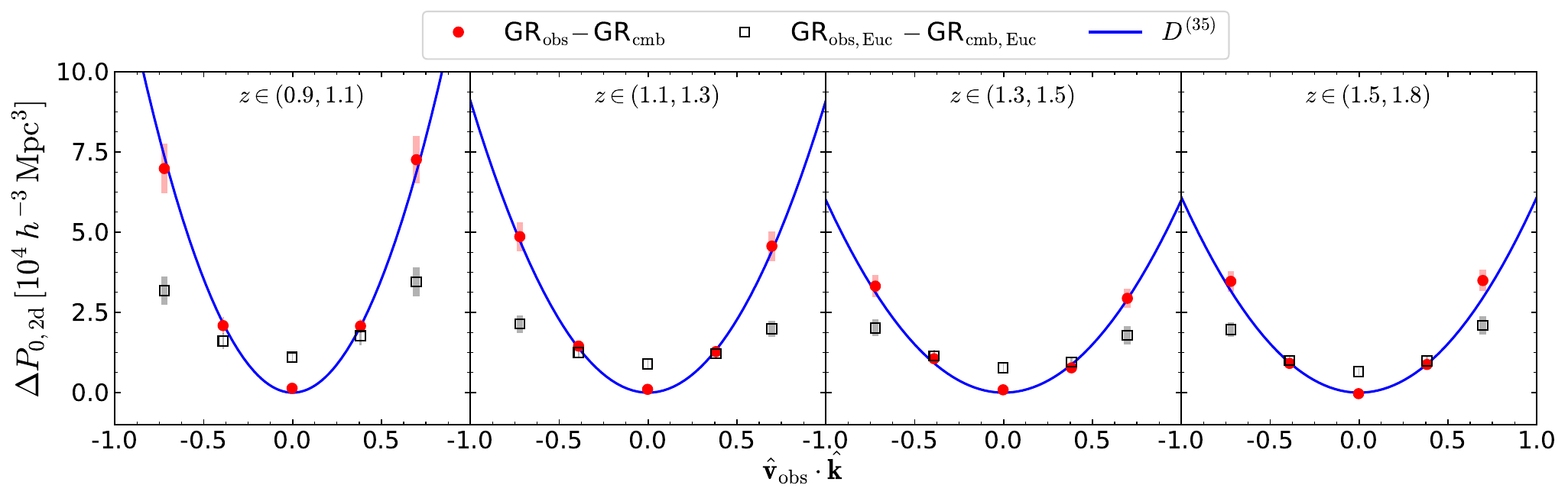}
    \caption{The difference between the two-dimensional monopole moments of the power spectra measured in the observer's and CMB rest frames as a function of $\hat{\mathbf{v}}_\text{obs}\cdot \hat{\mathbf{k}}$
    in the bin $k_\text{FFT}\leq k<6\,k_\text{FFT}$.
    The color scheme is as in Fig.~\ref{fig:j12} but this time the measurements from the simulations are represented by symbols with errorbars and the solid lines are computed using
    equation~(\ref{eq:meandipolespectrum}).}
    
    \label{fig:U2PLots}
\end{figure*}
The amplitude of the oscillations does not change much with redshift and
 the relativistic version
 of $\alpha(x)$ given in equation~(\ref{eq:dipoleobs}) 
perfectly explains the redshift dependence we see in the \textsc{liger} mocks. Since this function contains
terms that
depend on the expansion history of the Universe, it is intriguing to speculate that the
 large-scale oscillations in the monopole of the power spectrum measured in the observer frame
 could be used to 
 set constraints on the cosmological parameters. We will return to this matter below.

Going back to Fig.~\ref{fig:all_effects_euc}, we are 
now ready to explain why $P_0$ extracted from the V$_\text{obs}$ mocks lies well above its counterpart from the GR$_\text{obs}$ catalogues. This happens because the V$_\text{obs}$ mocks assume
$\mc{M}=1$ which corresponds to setting
$\mc{Q}=0$ in equation~(\ref{eq:alphaGRobs}) and thus
to a higher $\alpha$. On the other hand, the increase in the scatter between the different realisations is caused by the higher clustering signal and by the variability in $v_\text{obs}$.

In order to further 
investigate the clustering generated by the kinematic dipole overdensity,
we compute the monopole moment of the power spectrum 
by binning the Fourier modes as a function of both the wavenumber and the cosine of the angle between the wavevector and the peculiar velocity of the observer.
We refer to this quantity as the `two-dimensional' monopole moment of the power spectrum
and use the symbol $P_{0,2\text{d}}$ to denote it. Under the same assumptions as above, we thus have
\begin{align}
   \Delta P_{0, 2\text{d}}(k,\hat{\mathbf{k}}\cdot \hat{\mathbf{v}}_\text{obs})&=P_{0,2\text{d}}^{(\text{obs})}(k,\hat{\mathbf{k}}\cdot \hat{\mathbf{v}}_\text{obs})-P_{0,2\text{d}}^{(\text{cmb})}(k,\hat{\mathbf{k}}\cdot \hat{\mathbf{v}}_\text{obs})\nonumber \\
   &\simeq D(k,\hat{\mathbf{k}}\cdot \hat{\mathbf{v}}_\text{obs})\;.
   \end{align}
   Of course $P_{0,2\text{d}}^{(\text{cmb})}(k,\hat{\mathbf{k}}\cdot \hat{\mathbf{v}}_\text{obs})=P_0^{(\text{cmb})}(k)$ but the equality cannot be extended to the estimates from the individual mock catalogues due to the measurement noise.
In Fig.~\ref{fig:U2PLots}, we show
the dependence of $\Delta P_{0, 2\text{d}}$ on $\hat{\mathbf{v}}_\text{obs}\cdot \hat{\mathbf{k}}$ for the Fourier modes with $k_\text{FFT}\leq k<6\,k_\text{FFT}$. Symbols with errorbars denote the mean measurements from the mock catalogues while the solid lines represent  equation~(\ref{eq:meandipolespectrum}). The color coding is as in Fig.~\ref{fig:j12}. Once again the agreement between the model and the measurements is excellent for the full-sky mock catalogues that include all the relativistic effects.  On the other hand, similar to Fig.~\ref{fig:j12}, the Euclid-like results (empty squares) are affected by the window function which makes the angular dependence less prominent but generates a signal at $\hat{\mathbf{v}}_\text{obs}\cdot \hat{\mathbf{k}}=0$. 
\begin{figure*}

    \centering
	\includegraphics[width=1\textwidth]{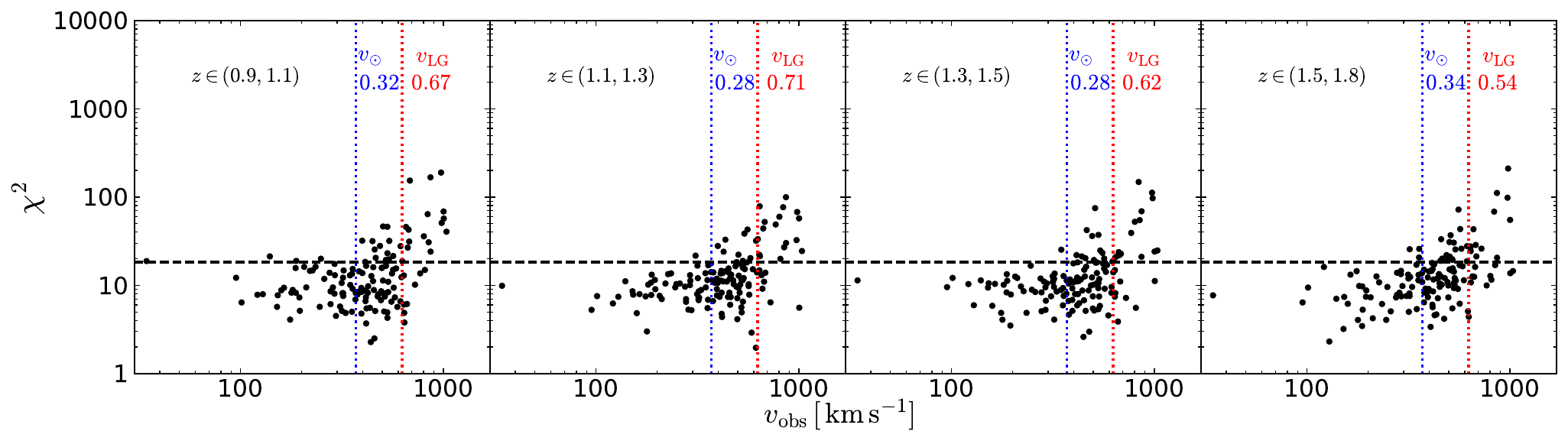}
    \caption{The $\chi^2$ goodness-of-fit statistic measured by fitting the mean $\langle \hat{P}_0^\text{(cmb)}\rangle$ for a Euclid-like survey
    to the individual measurements of $\hat{P}_0^\text{(obs)}$
    for each mock catalogue (corresponding to a different $v_\text{obs}$). The horizontal dashed line indicates the 95-per-cent confidence limit for 10 degrees of freedom. For the realisations that lie above this line, the difference between $\hat{P}_0^\text{(obs)}$ and the expected monopole in the CMB frame is statistically significant. The blue and red dotted lines represent the peculiar velocity of the solar system and the Local Group, respectively. Next to them we report the fraction of realisations with $v>v_\odot$ (or $v>v_\text{LG}$)  in which the $\chi^2$ is above the 95 per cent confidence limit.}
     \label{fig:CHI_VOBS}

\end{figure*}
\begin{figure}

    \centering
	\includegraphics[width=0.4\textwidth]{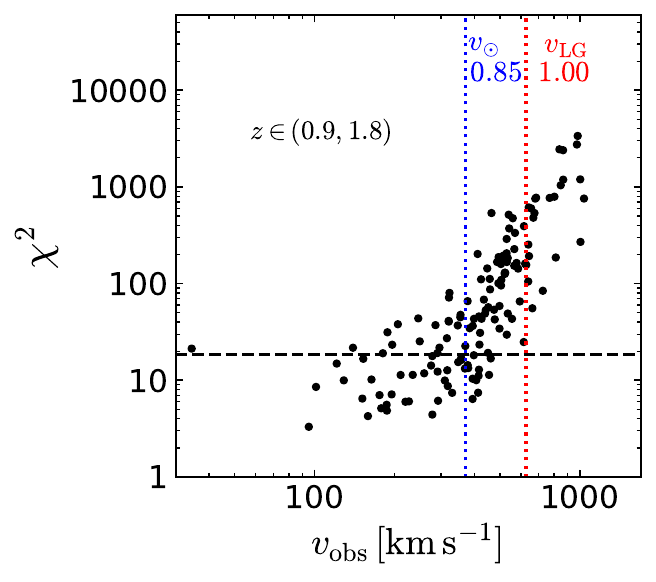}
    \caption{As in Fig.~\ref{fig:CHI_VOBS} but for a broader redshift bin.}
    
     \label{fig:CHI_VOBS_LARGE}

\end{figure}
So far we have concentrated on the averages over the 140 mock light cones.
In order to evaluate whether the impact of $\mathbf{v}_\text{obs}$ is detectable from a single survey, 
we compare $\hat{P}_0^\text{(obs)}$ extracted
from one of the GR$_\text{obs}$ mock catalogues to $\langle \hat{P}_0^\text{(cmb)}\rangle$
which includes all the relativistic
effect apart from those generated
by $\mathbf{v}_\text{obs}$. The idea is to check whether the model that
perfectly describes the monopole moment in the CMB frame
is ruled out by the data in the observer's frame
based on the $\chi^2$ goodness-of-fit statistic. In Figs.~\ref{fig:CHI_VOBS} and
\ref{fig:CHI_VOBS_LARGE}, we plot
the resulting values against
$v_\text{obs}$ for a Euclid-like survey. The horizontal dashed line
indicates the 95-per-cent confidence level for a $\chi^2$ distribution with 10 degrees of freedom. 
Whenever a point lies above this threshold, the model with no observer's effects is rejected
by the data at the selected confidence level. In other words,
the impact of $\mathbf{v}_\text{obs}$
should be
detectable in these realisations.
As expected, the detectability of the signal depends on $v_\text{obs}$ and
on the redshift interval under consideration.
For the broader redshift range considered in Fig.~\ref{fig:CHI_VOBS_LARGE}, the model in the CMB frame is ruled out in all the mocks 
with $v_\text{obs}>470$ km s$^{-1}$
and in most of those with
$v_\text{obs}\simeq 300$ km s$^{-1}$.

These results open up new opportunities for extracting cosmological information from galaxy surveys.
Usually, catalogues of extragalactic objects list `heliocentric'
redshifts obtained by correcting for Earth's rotational and orbital motions. In order to get redshifts in the CMB frame, one has to 
further account for the peculiar motion of the Sun. 
Under the assumption that the full CMB dipole is due to the Doppler effect, the velocity of the Sun with respect to the rest frame of the CMB is found to be $v_\odot= 369.82 \pm 0.11$  km s$^{-1}$ toward $(\ell, b) = (264.021\pm 0.011^\circ , 48.253 \pm 0.005^\circ)$
\citep{planck-dipole-18}.
Correcting this result for local motions,
it is possible to estimate the
peculiar velocity of the Local Group of galaxies. This gives
$v_\text{LG}=620\pm 15$ km s$^{-1}$ toward 
$(\ell, b) = (276^\circ, 30^\circ)$ with an uncertainty of a couple of degrees. 
{In Figs.~\ref{fig:CHI_VOBS} and \ref{fig:CHI_VOBS_LARGE}, $v_\odot$ and $v_\text{LG}$ are indicated with vertical blue and red lines, respectively.
Next to them, we also report the fraction
of realisation with a larger $v_\text{obs}$ in which the impact of the kinematic dipole overdensity is detectable at 95 per cent confidence.
The effect is strongly enhanced in the wide redshift bin} as seen in Fig.~\ref{fig:CHI_VOBS_LARGE}. In this case, $D_\text{iso}(k)$ in measurable in all the realizations with a velocity larger than $v_\text{LG}$ and in most of the realizations with $v_\text{obs}>v_\odot$.    

It is thus conceivable to fit the model for $D_\text{iso}$ (convolved with the window function of the survey) to the measurements of $\Delta P_0$ in order to determine the function $\alpha(x)$ (for instance, by constraining the coefficients of its Taylor expansion in a finite redshift range). If, at the same time, it was possible to precisely characterise the selection function of the survey and measure the faint-end slope of the galaxy luminosity function, knowing $\alpha(x)$ would provide information on the expansion rate of the Universe at the probed redshifts. Conversely, by assuming a cosmological model with the currently favoured values for the parameters to get $H(z)$, 
it would be possible to determine the magnification bias of the galaxy sample from  $\alpha(x)$.

Figs.~\ref{fig:CHI_VOBS} and \ref{fig:CHI_VOBS_LARGE} clearly show that dipoles generated by larger $v_\text{obs}$ would be more easily detectable with a Euclid-like survey. An interesting perspective then could be to 
artificially transform the galaxy redshifts to
a fictitious observer frame with a very large peculiar velocity. Although it appears as a free lunch, this should strongly enhance the dipole signal and thus allow
us to measure $\alpha(x)$ with a higher
signal-to-noise ratio. 

Finally, we want to speculate about the possibility of using 
$\Delta P_0$ and $\Delta P$ in order to determine the peculiar velocity of the Sun and thus test the assumption that the CMB dipole is fully kinematic. 
In this case, 
one would measure isotropic and anisotropic power spectra in the observer frame
directly from the heliocentric redshifts.
The redshift of a galaxy in the CMB frame is obtained using
\begin{equation}
    1+z_\text{cmb}
    =\frac{1+z_\text{obs}}
    {1+z_\text{pec, Sun}}
\end{equation}
where $z_\text{pec, Sun}\simeq \mathbf{v}_\odot\cdot \hat{\mathbf{x}}/c$ and $\hat{\mathbf{x}}$ the position vector of the galaxy. Here,
$\mathbf{v}_\odot$ should be treated as an unknown.
Basically, the idea would be to
determine $\mathbf{v}_\odot$ by identifying the peculiar-velocity vector such that the expressions in equations~(\ref{eq:dipolespectrum}) and $(\ref{eq:Diso})$ match
the measurements of the difference
between the spectra in the observer frame and in the putative CMB frame.
Since $D$ scales as a cosine squared,
this would lead to a degeneracy between vectors pointing in opposite directions which can be broken using CMB data.
Several challenges, however, undermine the success of this approach. First of all, it would require knowledge of the $\alpha$ function.
Moreover, it is important to keep in mind that only a small part (approximately one in a hundred) of the CMB dipole is expected to have an intrinsic origin. Thus, it is difficult to evaluate if this approach can
improve upon the current state of knowledge.
We will explore all these possibilities in our future work.

\section{Summary}\label{sec:summ}
The current standard model for RSD 
in the galaxy power spectrum \citep{Kaiser87}
relies upon the distant-observer
approximation and neglects light-cone
effects.
This reflects the fact that early galaxy redshift surveys covered small solid angles and were rather shallow in depth (relative to today's standards).
The current generation of instruments, however, allows us to survey major fractions of the sky
and measure spectroscopic redshifts
corresponding to comoving distances
which are comparable to the Hubble radius.

We first
address the pressing question
of whether the standard model of RSD need be modified 
to take into account wide-angle separations and general-relativistic light-cone effects. 
In order to compute the monopole moment of the observed galaxy power spectrum, $P_0$, without resorting to the plane-parallel approximation,
we use an upgraded implementation of the \textsc{liger} method which
we make publicly available at
this \href{https://astro.uni-bonn.de/\~porciani/LIGER/}{URL}.
\textsc{liger} combines low-resolution N-body simulations and a perturbative scheme to generate mock galaxy density fields in redshift space.
With its flexibility, \textsc{liger} allows us to quantify the impact of different effects by turning them on and off at will.
Moreover, by building many mock realisations of a survey, it also enable us to compute the covariance matrix
of the measurements (accounting for all effects and the window function of the survey) which is a challenging thing to do with analytic methods.

By considering a Euclid-like spectroscopic survey as a working example, we reach
the following conclusions.
\begin{enumerate}
    \item Wide-angle corrections due to peculiar velocities become important and cause a statistically significant increase of power for $k<0.02\,h$ Mpc$^{-1}$ (with respect to predictions made assuming the global plane-parallel approximation). Therefore, it is necessary to model them if one wants to consider these large scales for cosmological inferences. 
    \item The impact of other relativistic light-cone effects and, in particular, of gravitational lensing is small, provided that the current models for the luminosity function of H$\alpha$ emitters do not grossly underestimate the faint-end slope. Only for redshifts $z>1.5$ (or for very broad redshift bins), can magnification bias systematically shift the power-spectrum monopole by an amount which is comparable to the statistical uncertainties.
\end{enumerate}    

In the second part of this work, we investigate the impact of the peculiar velocity of the observer, $\mathbf{v}_\text{obs}$, on the monopole moment of the galaxy power spectrum. Our main findings are as follows.   
\begin{enumerate}   
  \setcounter{enumi}{2}
    \item 
    A non-vanishing $\mathbf{v}_\text{obs}$
    alters the redshifts of the galaxies and generates a dipolar pattern in the reconstructed galaxy distribution in redshift space that we name the finger of the observer effect. In consequence,
    the monopole moment of the power spectrum measured in the observer's rest frame has an additive component showing a characteristic damped oscillatory pattern on large scales. We derive an analytical
    expression for this term which exquisitely fits our numerical results.
    \item The amplitude of the oscillating correction, $\Delta P_0$, obviously depends on $v_\text{obs}$. Given the current estimate of the peculiar velocity of the Sun, we find that the $\Delta P_0$ signal should be detectable above the noise in a Euclid-like survey, particularly when one considers broad redshift bins.
    \item The function $\Delta P_0$
     also depends on several properties of the galaxy population under study and the expansion rate of the Universe.
    By comparing the galaxy power spectra measured in the CMB and observer's rest frames, it is thus possible to make inferences about these quantities. This opens up interesting perspectives that we will explore in our future work.
\end{enumerate}

\section*{Acknowledgements}
CP thanks Emanuele Castorina for discussions. DB acknowledges partial financial support by ASI Grant No. 2016-24-H.0. MYE and DB acknowledge funding from Italian Ministry of Education, University and Research (MIUR) through the ``Dipartimenti di eccellenza" project ``Science of the Universe".

The data analysis was performed
using \ts{numpy} \citep{numpy} , \ts{scipy} \citep{scipy} and \ts{pandas} \citep{pandas}. The plots were created using using \ts{matplotlib} \citep{matplotlib} and \ts{cmasher} \citep{CMASHER}.

While we were completing this work, we became aware of a few pre-prints
by Nadolny et al. (\href{https://arxiv.org/abs/2106.05284}{arXiv:2106.05284}),
Castorina \& di Dio (\href{https://arxiv.org/abs/2106.08857}{arXiv:2106.08857}),
and Kalus et al. (\href{https://arxiv.org/abs/2107.00351}{arXiv:2107.00351})
which are complementary to our results on the impact of $\mathbf{v}_\text{obs}$.
\section*{Data Availability}
The data underlying this article will be shared on reasonable request to the corresponding author.



\bibliographystyle{mnras}
\bibliography{example} 




\appendix
\section{LIGER II}
We briefly summarise here
the basics of the \ts{liger} method and also describe the improvements
that are implemented in the new version of the code.
\subsection{Particle shifts}\label{app:shift_eqs}
The \ts{liger} method computes
a 4-dimensional coordinate transformation between the
real-space position $x^{\mu}_{\rm{r}}$ and the redshift-space position $x^{\mu}_{\rm{s}}$ of the simulation particles. This is obtained by studying
the propagation of a light beam from an emitter to an observer in a perturbed
FLRW model.
Following the notation of Paper I, we write
\begin{equation}\label{eq:coor_map}
 x^{\mu}_{\rm{r}} = x^{\mu}_{\rm{s}} + \Delta x^{\mu}\,,
\end{equation}
with
\begin{align}
\label{eq:shift_lig}
    \Delta x^{0} &= \frac{1}{\mc{H}}\,\delta \ln a\,, \\
    \Delta x^{i} &= -\chi_{\rm{s}}\left[ n^i_{\rm{s}}\,(\Phi_{\rm{o}}+\Psi_{\rm{o}} )+ v^i_{\rm{o}} -n^i_{\rm{s}}\;(n^j_{\rm{s}}v_j)_{\rm{o}}  \right] - \frac{1}{\mc{H}} \delta \ln a \nonumber \\ \nonumber
    & + \int^{\chi_{\rm{s}}}_0 (\chi_{\rm{s}} - \chi)\left[n^i_{\rm{s}}\partial_0(\Phi+\Psi)-\delta^{i}_j\partial^j(\Phi+\Psi)\right]\\
    &+ 2 n^i_{\rm{s}} \int^{\chi_{\rm{s}}}_0 (\Phi+\Psi) \,\dif \chi\,,
\end{align}
where
\begin{equation}
    \delta \ln a = \Phi_{\rm{o}} - (n^j_{\rm{s}}v_j)_{\rm{o}} - \Phi_{\rm{e}} +(n^j_{\rm{s}}v_j)_{\rm{e}} - \int^{\chi_{\rm{s}}}_0 \partial_0(\Phi+\Psi)\, \dif \chi\,,
\end{equation}
quantifies the apparent redshift change -- i.e. $\delta \ln a=\delta z/(1+z)$ -- due to the perturbations at every point on the past light cone of the observer,
$n^i_s$ denotes the observed direction of the source,  $v^i$ is the peculiar velocity, $\mc{H} = \partial_0 \ln a$, $\Psi$ and $\Phi $ are the Bardeen potentials and
the subscripts `e' and `o' denote functions evaluated at the locations of the light source and of the observer, respectively.
Note that the transformation above
goes from the redshift-space position to the real-space one while in the code 
we implement the inverse transformation
(see section 2.1 of Paper I for details).
Similarly, the magnification of the source (lensing+Doppler) is computed using
\begin{align}
\label{eq:Mag_lig}
\mc{M} &= 1 + 2\Phi_\text{e} - \,2\left(1-\frac{1}{\mc{H}\chi_{\rm{s}}} \right)\,\delta \ln a - 2\,(n^j_{\rm{s}}v_j)_{\rm{o}} \\\nonumber
&+ \int^{\chi_{\rm{s}}}_0 (\chi_{\rm{s}} - \chi) \frac{\chi}{\chi_{\rm{s}}} \nabla_{\bot} (\Phi+\Psi) \dif \, \chi ^2\\\nonumber
&- \frac{2}{\chi_{\rm{s}}}\int^{\chi_{\rm{s}}}_0 (\Phi+\Psi)\, \dif\, \chi\,\,.
\end{align}
The current implementation assumes
a $\Lambda$CDM model where $\Phi=\Psi$.

As explained in the main text, in this work, we consider five sets of mock catalogues which are given different names to distinguish them.\begin{enumerate}
\item In the REAL mocks, we set $\mc{M}=1$ and $\Delta x^\mu=0$ (no RSD).
\item
Those dubbed V$_\text{cmb}$ 
are obtained by setting $\mc{M}=1$ and 
only considering the term
$(n^j_{\rm{s}}v_j)_{\rm{e}}$ in
$\Delta x^\mu$.
\item The
GR$_\text{cmb}$ mocks are derived using the equations
for $\Delta x^\mu$ and $\mc{M}$ given above
with $v^i_\text{o}=\Phi_\text{o}=\Psi_\text{o}=0$.
\item We build the V$_\text{obs}$ mocks by setting $\mc{M}=1$ and 
only considering the terms
$(n^j_{\rm{s}}v_j)_{\text{e}}$ and
$(n^j_{\rm{s}}v_j)_{\text{o}}$ which appear in $\delta \ln a$
to compute $\Delta x^\mu$.
\item Finally, the GR$_\text{obs}$ mocks include all the terms.
\end{enumerate}

\subsection{Galaxy density field}
\label{galdens}

In the continuum approximation and to linear order in the perturbations, we can express
the galaxy density contrast in redshift space as (see section 2.3 in Paper I)
\begin{equation}
    \delta_{\text{g}}\simeq
    (b-1) \,\delta_{\mathrm{r}}+\delta_{\mathrm s}+\mathcal{E}\, \delta \ln a+\mathcal{Q}\,(\mathcal{M}-1)\;.
    \label{eq:master_eq}
\end{equation}
Note that
$b, \mathcal{E}$ and $\mathcal{Q}$ are functions of $z$ only, while the density contrasts, $\delta \ln a$ and $\mathcal{M}$
depend on both $\hat{\mathbf{n}}_\mathrm{s}$ and $z$.

In order to estimate the matter density contrast 
starting from the particle positions in our N-body simulations, we use a standard mass-weighted smoothing scheme,
\begin{align}
   1+\delta(\mathbf{x})&= \frac{\eta(\mathbf{x})}{\bar{\eta}}=
   \frac{1}{\bar{\eta}}\,\int \eta_{\mathrm{part}}(\mathbf{y})\,W_\text{CIC}(\mathbf{x}-\mathbf{y})\,\dif^3y\nonumber\\
   &=\frac{1}{\bar{\eta}}\,\int \left[\sum_i \delta_\mathrm{D}(\mathbf{y}_i-\mathbf{y})\right]\,W_\text{CIC}(\mathbf{x}-\mathbf{y})\,\dif^3y \nonumber\\
&=   
   \frac{1}{\bar{\eta}}\,\sum_i W_\text{CIC}(\mathbf{x}-\mathbf{y}_i)\;,
\end{align}
where $\eta$, $\eta_{\mathrm{part}}$ and $\bar{\eta}$ are the smooth, discrete and average particle number densities, respectively, the index $i$ runs over the simulation particles, and $W_\text{CIC}$ denotes the `cloud-in-cell' kernel \citep{Hockney-Eastwood}.
We use the real-space particle positions to get $\delta_\mathrm{r}$ and the redshift-space ones to get $\delta_\mathrm{s}$. We sample the continuous fields $\delta_\mathrm{r}$ and $\delta_\mathrm{s}$ on the same regular Cartesian grid that covers the past light cone of the observer.

The estimation of the fields $\delta \ln a$ and $\mathcal{M}$ involves additional complications due to the fact that we only know the redshift change and the lensing magnification at the positions of the simulation particles. In other words, we have a better spatial sampling of these quantities where the matter density is high and a poorer sampling where it is low. This forces us to 
use mass-weighted averages instead of volume-weighted ones. 
Therefore, we write
\begin{align}
   \mathcal{M}(\mathbf{x})&=
   \frac{\int \mathcal{M}_{\mathrm{part}}(\mathbf{y})\, \eta_{\mathrm{part}}(\mathbf{y})\,W_\text{CIC}(\mathbf{x}-\mathbf{y})\,\dif^3y}
   {\int \eta_{\mathrm{part}}(\mathbf{y})\,W_\text{CIC}(\mathbf{x}-\mathbf{y})\,\dif^3y}\nonumber \\
   &=\frac{\displaystyle{\sum_i} \mathcal{M}_{\mathrm{part}}(\mathbf{y}_i)\,W_\text{CIC}(\mathbf{x}-\mathbf{y}_i)}
   {\displaystyle{\sum_i} W_\text{CIC}(\mathbf{x}-\mathbf{y}_i)}\;,
\end{align}
and a corresponding expression for $\delta \ln a$. 
Our estimators are unbiased at linear perturbative order but deviate from the actual fields at second order. In fact,
since multiplication and smoothing do not commute, 
$\overline{\mathcal{M}}\simeq \overline{\mathcal{M}_1}+\overline{\mathcal{M}_2}+\overline{\mathcal{M}_1\,\delta_1}-\overline{\mathcal{M}_1}\,\overline{\delta_1}+\dots$
where -- just for this expression -- we have denoted smoothed quantities
with an overbar and the indices indicate the order of the different perturbative contributions.
It is worth mentioning that the
same method is often used to estimate the peculiar velocity field in N-body simulations \citep[e.g.][]{Bernardeau+vandeweigaert95}.

Eventually, we build the galaxy distribution on the past light cone of the observer by writing $n_{\mathrm{ g}}(\mathbf{x})=\bar{n}_\mathrm{g}(z)\,[1+\delta_{\mathrm{g}}(\mathbf{x})]$.
   
\subsection{Shot noise}
\label{sec:shotnoise}
Finally, in order to add (Poissonian) shot noise to the \textsc{liger} mocks, we replace the expected number of galaxies in each cell, $N_\text{g}(\mathbf{x})=\bar{n}_\text{g}(z)\,[1+\delta_\text{g}(\mathbf{x})]\,V_\mathrm{cell}$,
with a random deviate extracted from a Poisson distribution with mean $N_\text{g}(\mathbf{x})$.

\section{Evolution and magnification  bias}
\label{sec:evmagbias}
 Let us denote by $\phi(L,z)$ 
 the (background) galaxy luminosity function defined
such that the comoving number density of galaxies brighter than the threshold $L_\text{min}$ at redshift $z$ is
\begin{equation}
    n(L_\text{min},z)=\int_{L_\text{min}}^\infty 
    \phi(L,z)\,\dif L\;.
\end{equation}
The number density of galaxies that can be detected above the flux limit $f_\text{lim}$ by an observer at $z=0$ is
\begin{equation}
    \bar{n}_\text{g}(z)=n(L_\text{lim}(z),z)=\int_{L_\text{lim}(z)}^\infty 
    \phi(L,z)\,\dif L\;,
\end{equation}
with $L_\text{lim}(z)=4 \pi\, D_\mathrm{L}^2(z)\, f_\text{lim}$ 
and $D_\mathrm{L}(z)$ the luminosity distance.
In order to account for the fact that
a redshift survey does not successfully measure the redshift of all galaxies, the observed number counts
need to be multiplied by the completeness fraction $0 \leq c_z \leq1$ that we assume to be
a constant, for simplicity.

The evolution and magnification biases are defined in equations~(\ref{eq:evbias}) and (\ref{eq:magbias}) as the partial logarithmic derivatives of $n$ evaluated at $L_\text{min}=L_\text{lim}(z)$.
Note that, being logarithmic derivatives, both $\mc{Q}$ and $\mc{E}$ do not depend on $c_z$.
It follows from the definitions that
\begin{align}
\label{eq:Q_bias}
    \mc{Q}(z)&= -\frac{L_\text{min}}{n(L_\text{min},z)}\,\left. \frac{\partial  n(L_\text{min},z)}{\partial L_\text{min}}\right|_{L_\text{min}=L_\text{lim}(z)}\nonumber \\
    &=\frac{L_\text{lim}(z)\,\phi(L_\text{lim}(z),z)}{\bar{n}_\text{g}(z)}\;.
\end{align}
If the flux limit probes the faint-end of the luminosity function where $\phi\propto L^{-\gamma}$ with $\gamma>1$,
we obtain $\mc{Q}\simeq \gamma-1$.
In simple words, $\mc{Q}$ reflects
the slope of the (unlensed) galaxy luminosity function at the value of $L$ corresponding to the flux limit of the survey.
Similarly, for the evolution bias,
we obtain
\begin{align}
\mc{E}(z)&=
-\left.\frac{\partial \ln \int_{L_\text{min}}^\infty 
    \phi(L,z)\,\dif L}{\partial \ln (1+z)}  
\,
\right|_{L_\text{min}=L_\text{lim}(z)}\nonumber\\
&=-\frac{1}{\bar{n}_\text{g}(z)} \,\int_{L_\text{lim}(z)}^\infty 
    \frac{\partial \phi(L,z)}{\partial \ln (1+z)}  \,\dif L  \;. 
\end{align}
Therefore $\mc{E}$ quantifies 
how the amplitude and the shape of the luminosity function 
change with $z$ for $L>L_\text{lim}(z)$.

It is interesting to relate $\mc{E}(z)$
to the logarithmic derivative of $\bar{n}_\text{g}$, with which it is often confused. We have 
\begin{align}
&-\frac{\dif \ln \bar{n}_\text{g}(z)}{\dif \ln (1+z)}
=-\frac{1}{\bar{n}_\text{g}(z)}\,\frac{\dif  \bar{n}_\text{g}(z)}{\dif \ln (1+z)} \nonumber\\
&=-\frac{1}{\bar{n}_\text{g}(z)}\,\frac{\dif  \int_{L_\text{lim}(z)}^\infty 
    \phi(L,z)\,\dif L}{\dif \ln (1+z)}\nonumber \\
&=-\frac{1}{\bar{n}_\text{g}(z)}\,\left[- \phi(L_\text{lim}(z),z)\,\frac{\dif L_\text{lim}(z)}{\dif \ln (1+z)}+\int_{L_\text{lim}(z)}^\infty 
    \frac{\partial \phi(L,z)}{\partial \ln (1+z)}  \,\dif L\right]\nonumber \\
  &=\frac{L_\text{lim}(z)\,\phi(L_\text{lim}(z),z)}{\bar{n}_\text{g}(z)}\,\frac{\dif \ln L_\text{lim}(z)}{\dif \ln (1+z)}+\mc{E}(z) \nonumber \\
  &=\mc{Q}(z)\,\frac{\dif \ln L_\text{lim}(z)}{\dif \ln (1+z)}+\mc{E}(z) \nonumber \\
  &=2\,\mc{Q}(z)\,\frac{\dif \ln D_\text{L}(z)}{\dif \ln (1+z)}+\mc{E}(z)\;.
\end{align}
In a flat universe,
$D_\text{L}(z)=(1+z)\, x(z)$ and
\begin{align}
    \frac{\dif D_\text{L}(z)}{\dif  (1+z)}
=x(z)+\frac{c\,(1+z)}{H(z)}\;.
\end{align}
Putting everything together, we get \citep[see also][]{Bertacca+15}
\begin{align}
 -\frac{\dif \ln \bar{n}_\text{g}(z)}{\dif \ln (1+z)}=
 2\,\mc{Q}(z)\,\left[1+ \frac{c\,(1+z)}{H(z)\,x(z)} \right]+\mc{E}(z)
 \end{align}
which is equivalent to equation~(\ref{eq:alpha_c}).
\section{Angular power spectra} 
\label{ligcell}
In order to estimate the angular power spectra from the \ts{liger} mocks, we create digitized full-sky maps of the projected
density contrast using a Hierarchical Equal Area isoLatitude Pixelation \citep[HEALPix,][]{HEALpix} of the celestial sphere with $N_{\mathrm{pix}}=12\times(1024)^2$ pixels. For each area element on the sky, we measure the projected number-density contrast 

using the line-of-sight integral \citep{C_LEST} 
\begin{equation}\
\label{C_L_INTEGRAL}
\Sigma(\Omega)\, = \, \frac{\int_0^{\infty}W(\mathbf{x})\, n_g(\mathbf{x})\,x^2 \,\dif x}{\int_0^{\infty} W(\mathbf{x})\,\widehat{n}(x)\,x^2\, \dif x} -1\;,
\end{equation}
where $W(\mathbf{x})$ denotes a generic window function for which we adopt a top-hat function between a minimum and
a maximum redshift.
In order to compute the integral,
we implement the fast voxel traversal algorithm for ray tracing introduced by \citet{RAY}.
By taking advantage of the HEALPix software package,
we decompose the maps into spherical harmonics
\begin{equation}
a_{\ell m}=\int  \Sigma(\Omega)\, Y^*_{\ell m}(\Omega) \,\dif\Omega\,,
\end{equation}
and measure the angular power spectra using 
\begin{equation}
\hat{C}_{\ell}=\frac{1}{w_\text{p}^{2}\,(2\ell + 1)} \sum_{m=-\ell}^\ell |a_{\ell m}|^2 \,-\frac{4\pi}{N} \,,
\end{equation}
where $w_\text{p}$ is a correction factor of  order unity due to the finite resolution of the HEALPix maps
(see \url{https://healpix.jpl.nasa.gov/html/intronode14.htm}).

\bsp	
\label{lastpage}
\end{document}